\documentclass[10pt]{amsart}

\usepackage{lipsum}
\usepackage{amsfonts,bm,amssymb,amsmath}
\usepackage{graphicx}
\usepackage{epstopdf}
\usepackage[caption=false]{subfig}
\usepackage{pgfplots}
\usepackage{algorithmic}
\usepackage{amsopn}
\usepackage{amsrefs}
\usepackage{float}
\usepackage{caption}

\usepackage{hyperref}
\usepackage{cleveref}

\usepackage{color}
\usepackage{diagbox}

\usepackage{geometry}

\hypersetup{
	colorlinks,
	linkcolor={red!50!black},
	citecolor={blue!50!black},
	urlcolor={blue!80!black}
}

\numberwithin{equation}{section}

\definecolor{newcolor1}{rgb}{.8,.349,.1}
\colorlet{bblue}{blue!50!black}
\crefformat{equation}{(#2#1#3)}
\crefmultiformat{equation}{(#2#1#3)}{ and~(#2#1#3)}{, (#2#1#3)}{ and~(#2#1#3)}

\crefformat{figure}{Figure~#2#1#3}
\crefmultiformat{figure}{Figures~ #2#1#3}{ and~#2#1#3}{, (#2#1#3)}{ and~(#2#1#3)}
\crefformat{table}{Table~#2#1#3}

\def\a{\mbox{\boldmath $a$}}

\def\e{\mbox{\boldmath $e$}}
\def\f{\mbox{\boldmath $f$}}

\def\g{\mbox{\boldmath $g$}}
\def\h{\mbox{\boldmath $h$}}
\def\m{\mbox{\boldmath $m$}}

\def\x{\mbox{\boldmath $x$}}
\def\y{\mbox{\boldmath $y$}}

\def\0{\mbox{\boldmath $0$}}

\begin{document}

\title[A third-order method for micromagnetics]{Enhancing Micromagnetics Simulations with a Third-Order Semi-Implicit Projection Method}

\author[C. Xie]{Changjian Xie}
\address{School of Mathematics and Physics\\ Xi'an-Jiaotong-Liverpool University\\Re'ai Rd. 111, Suzhou, 215123, Jiangsu\\ China.}
\email{Changjian.Xie@xjtlu.edu.cn}

\author[C. Wang]{Cheng Wang}
\address{Mathematics Department\\ University of Massachusetts\\ North Dartmouth\\ MA 02747\\ USA.}
\email{cwang1@umassd.edu}


\date{\today}


\begin{abstract}

Micromagnetics depends on high-fidelity numerical methods for magnetization dynamics. This work proposes a third-order temporal accuracy scheme for the Landau-Lifshitz-Gilbert equation, addressing accuracy-efficiency trade-offs in existing methods. Validated via nanostrip simulations (representative of real devices), the scheme offers two key advantages: rigorous third-order accuracy (surpassing existing simulation methods) and higher computational efficiency, ensuring fast convergence without precision loss. It maintains stability for Gilbert damping \(\alpha\) from $0.1$ to $10$, avoiding non-physical states. The magnetic microstructures it captures are consistent with established methods, confirming reliability for physical analysis.
\end{abstract}

\maketitle

\section{Introduction}

Ferromagnetic materials have found widespread application in data storage systems, owing to the bi-stable characteristics of their intrinsic magnetic order—commonly referred to as magnetization. The dynamic evolution of magnetization in such materials is predominantly described by the Landau-Lifshitz-Gilbert (LLG) equation \cite{Landau1935On,Gilbert:1955}. Notably, the LLG equation encompasses two fundamental terms that govern magnetization dynamics: the gyromagnetic term, which maintains energy conservation, and the damping term, which accounts for energy dissipation.
The damping term plays a pivotal role in magnetic devices, as it exerts a significant influence on both the energy consumption and operational speed of such systems. A recent experimental study conducted on magnetic-semiconductor heterostructures \cite{Zhang2020ExtremelyLM} has confirmed the adjustability of the Gilbert damping constant. At the microscopic scale, the damping effect arises from multiple mechanisms, including electron scattering, itinerant electron relaxation \cite{Heinrich1967TheIO}, and phonon-magnon coupling \cite{Suhl1998TheoryOT, Nan2020ElectricfieldCO}; these mechanisms can be quantified through electronic structure calculations \cite{TangXia2017}. From a practical application standpoint, tuning the damping parameter enables the optimization of magnetodynamic properties in ferromagnetic materials—for example, reducing the switching current and enhancing the writing speed of magnetic memory devices \cite{Wei2012MicromagneticsAR}.
While the majority of existing experimental research has focused on scenarios involving small damping parameters \cite{Budhathoki2020LowGD,Lattery2018LowGD,Weber2019GilbertDO}, pronounced large damping effects have been reported in studies such as \cite{GilbertKelly1955, Tanaka2014MicrowaveAssistedMR}. Specifically, Tanaka et al. \cite{Tanaka2014MicrowaveAssistedMR} observed that a large damping constant results in a shorter magnetization switching time. Additionally, Gilbert and Kelly \cite{GilbertKelly1955} documented the existence of extremely large damping parameters (approximately 9) in their work.

The LLG equation is a vectorial, nonlinear mathematical system characterized by the point-wise constant magnitude constraint of the magnetization vector. Substantial research efforts have been dedicated to developing efficient and numerically stable methods for micromagnetic simulations; comprehensive reviews and relevant references can be found in \cite{kruzik2006recent,cimrak2007survey}. Among the existing numerical approaches, semi-implicit schemes have gained considerable popularity due to their ability to circumvent complex nonlinear solvers while ensuring favorable numerical stability \cite{alouges2006convergence, gao2014optimal, Xie2018}.
For instance, We have proposed a second-order accurate backward differentiation formula (BDF2) scheme based on one-sided interpolation by \cite{Xie2018}. Such a scheme requires solving a three-dimensional linear system with non-constant coefficients at each time step. Furthermore, we have established a rigorous theoretical framework to derive the second-order convergence rate of this BDF2 method by \cite{jingrun2019analysis}. As an alternative approach, Alouges et al. \cite{alouges2006convergence} developed a linearly implicit method that employs the tangent space to satisfy the magnetization magnitude constraint, though this method only achieves first-order temporal accuracy.
In a more recent advancement, Lubich et al. \cite{Lubich2021} constructed and analyzed high-order BDF schemes for LLG equation. Notably, the unconditional unique solvability of semi-implicit schemes has been proven in \cite{jingrun2019analysis,Lubich2021}; however, their convergence analysis relies on the constraint that the temporal step size is proportional to the spatial grid size. Despite these progresses, the realistic simulation of high-order numerical methods for micromagnetic modeling remains an active research focus.
Nevertheless, existing numerical approaches still suffer from inherent limitations: most of them only achieve first-order or second-order temporal accuracy. Moreover, third-order accurate methods specifically tailored for practical micromagnetic simulations—especially those capable of handling arbitrary damping parameters and facilitating quantitative comparisons between different numerical methods—are currently lacking. To address this critical gap, this paper proposes a third-order accurate numerical method for solving the LLG equation with arbitrary damping parameters.
Additionally, extensive computational experiments have been carried out to validate the numerical stability of the proposed method. These experiments further confirm that the dissipation characteristics of the new scheme differ from those of our previously developed method \cite{xie2025schemeB} under conditions involving large damping parameters.

The structure of the rest of this paper is organized as follows. In \cref{sec: numerical scheme}, we first review the micromagnetic model, then elaborate on the proposed numerical method, and conduct a comparative analysis with the first-order (BDF1) and second-order (BDF2) semi-implicit projection methods. In \cref{sec:experiments}, we present comprehensive numerical results, including verifications of temporal and spatial accuracy in both one-dimensional (1D) and three-dimensional (3D) simulations, investigations into computational efficiency (via comparisons with the BDF1 and BDF2 algorithms), stability analyses with respect to the damping parameter, and explorations of the dependence of domain wall velocity on both the damping parameter and external magnetic field. Finally, concluding remarks and future research directions are provided in \cref{sec:conclusions}.

\section{The physical model and the numerical scheme}
\label{sec: numerical scheme}

\subsection{Governing equation}

The Landau-Lifshitz-Gilbert (LLG) equation constitutes the cornerstone of micromagnetics, rigorously describing the spatiotemporal dynamics of magnetization in ferromagnetic materials by encapsulating two fundamental physical processes: gyromagnetic precession and dissipative relaxation \cite{Landau1935On,Brown1963micromagnetics}. In its nondimensionalized form, this governing equation is mathematically formulated as
\begin{align}\label{c1-large}
{\m}_t =-{\m}\times{\bm h}_{\text{eff}}-\alpha{\m}\times({\m}\times{\bm h}_{\text{eff}})
\end{align}
subject to the homogeneous Neumann boundary condition
\begin{equation}\label{boundary-large}
\frac{\partial{\m}}{\partial {\bm \nu}}\Big|_{\partial \Omega}=0,
\end{equation}
where \(\Omega \subset \mathbb{R}^d\) (\(d=1,2,3\)) denotes the bounded domain occupied by the ferromagnetic specimen, and \(\bm \nu\) represents the unit outward normal vector on the domain boundary \(\partial \Omega\). This boundary condition enforces the absence of magnetic charge accumulation on the material surface, a physically consistent constraint for isolated ferromagnetic systems.

The magnetization field \(\m: \Omega \to \mathbb{R}^3\) is a three-dimensional vector field that adheres to the intrinsic pointwise constraint \(|\m| = 1\), a direct consequence of the quantum mechanical alignment of electron spins in ferromagnetic materials. The first term on the right-hand side (RHS) of \cref{c1-large} characterizes the gyromagnetic precession, wherein magnetic moments precess about the effective field \(\bm h_{\text{eff}}\) with angular frequency proportional to the gyromagnetic ratio. The second term quantifies dissipative relaxation, with \(\alpha > 0\) denoting the dimensionless Gilbert damping coefficient, which regulates the rate of energy dissipation into the material lattice.

From the perspective of the Gibbs free energy functional

The effective magnetic field \(\bm h_{\text{eff}}\) is derived from the functional derivative of the Gibbs free energy functional \(F[\m]\) with respect to the magnetization field, i.e., \(\bm h_{\text{eff}} = -\delta F/\delta \m\). This functional encapsulates all energy contributions inherent to ferromagnetic systems, including exchange, anisotropy, magnetostatic (stray), and Zeeman energies, and is expressed as
\begin{equation}\label{LL-Energy}
F[\m] = \frac {\mu_0 M_s^2}{2} \left\{\int_\Omega \left( \epsilon|\nabla\m|^2 +
q\left(m_2^2 + m_3^2\right)
-2\h_e\cdot\m - \h_s\cdot\m \right)\mathrm{d}\x \right\} . 
\end{equation}
wherein \(\mu_0 = 4\pi \times 10^{-7}\, \text{H/m}\) is the vacuum permeability, \(M_s\) denotes the saturation magnetization, and \(\epsilon\) and \(q\) are dimensionless parameters defined below. The vector fields \(\h_e\) and \(\h_s\) represent the external applied magnetic field and the stray field, respectively.
For uniaxial ferromagnetic materials—exhibiting a single crystallographic axis of easy magnetization, say the effective field \(\bm h_{\text{eff}}\) decomposes into physically distinct components, yielding the explicit form
\begin{align}
{\bm h}_{\text{eff}} =\epsilon\Delta\m-q(m_2\e_2+m_3\e_3)+\h_s+\h_e,
\end{align}
where \(\epsilon = C_{\text{ex}}/(\mu_0 M_s^2 L^2)\) and \(q = K_u/(\mu_0 M_s^2)\). Here, \(L\) is the characteristic length scale of the specimen, \(C_{\text{ex}}\) is the exchange constant (governing short-range spin alignment), and \(K_u\) is the uniaxial anisotropy constant (quantifying the energy penalty for magnetization misalignment with the easy axis). The unit vectors \(\e_2 = (0,1,0)\) and \(\e_3 = (0,0,1)\) define the hard axes transverse to the uniaxial easy axis, and \(\Delta\) denotes the Laplacian operator in \(d\)-dimensional space.
For Permalloy (NiFe), a prototypical soft ferromagnetic material widely utilized in spintronic devices, the following material parameters are well-established in the literature: \(C_{\text{ex}} = 1.3 \times 10^{-11}\, \text{J/m}\), \(K_u = 100\, \text{J/m}^3\), and \(M_s = 8.0 \times 10^5\, \text{A/m}\). The stray field \(\h_s\) arises from magnetic charge distributions at domain boundaries and material surfaces, and is mathematically described by the integral equation
\begin{align}\label{eqn:div}
	{\h}_{\text{s}}=\frac{1}{4\pi}\nabla \int_{\Omega} \nabla\left( \frac{1}{|\x-\y|}\right)\cdot {\bm m}({\bm y})\,d{\bm y},
\end{align}
which is a formulation that exhibits long-range spatial correlations. A critical computational advancement for practical micromagnetic simulations is that for rectangular domains \(\Omega\), the evaluation of \(\h_s\) can be efficiently computed via the Fast Fourier Transform (FFT) \cite{Wang2000}, which reduces the asymptotic computational complexity from \(O(N^d)\) to \(O(N^d \log N)\) for \(d\)-dimensional grids, enabling large-scale simulations.

To facilitate numerical discretization, we introduce the composite source term
\begin{align}\label{eq-4}
\f=-Q(m_2\e_2+m_3\e_3)+\h_s+\h_e.
\end{align}
which aggregates the anisotropy, stray field, and external field contributions. Substituting this source term into \cref{c1-large}, the LLG equation is re-expressed as
\begin{align}\label{eq-5}
\m_t=-\m\times(\epsilon\Delta\m+\f)-\alpha\m\times\m\times(\epsilon\Delta\m+\f).
\end{align}
Leveraging the vector triple product identity \(\a \times ({\bm b} \times {\bm c}) = (\a \cdot {\bm c}){\bm b} - (\a \cdot {\bm b}){\bm c}\) and the pointwise constraint \(|\m| = 1\) (which implies \(\m \cdot \partial_t \m = 0\) via time differentiation), we simplify \cref{eq-5} to an equivalent formulation that is more amenable to stable numerical discretization:
\begin{equation}\label{eq-model}
\m_t=\alpha  (\epsilon\Delta\m+\f)+\alpha \left(\epsilon |\nabla \m|^2 -\m \cdot\f \right)\m-\m\times(\epsilon\Delta\m+\f).
\end{equation}
We establish a standardized discretization framework to underpin subsequent numerical approximations, defining temporal and spatial discretization notations and boundary condition enforcement strategies.

Let \(k = t^{n+1} - t^n\) denote the uniform temporal step-size, with discrete time levels given by \(t^n = nk\) for \(n = 0,1,\dots,N_T\), where \(N_T = \lfloor T/k \rfloor\) and \(T\) is the final simulation time. For spatial discretization, a uniform Cartesian grid is employed with mesh-size \(h_x = h_y = h_z = h = L/N\) (for cubic domains of characteristic length \(L\)). The notation \(\m_{i,j,\ell}^n\) denotes the numerical approximation of \(\m(x_{i-1/2}, y_{j-1/2}, z_{\ell-1/2}, t^n)\), where \(x_{i-1/2} = (i - 1/2)h_x\), \(y_{j-1/2} = (j - 1/2)h_y\), and \(z_{\ell-1/2} = (\ell - 1/2)h_z\) define cell-centered grid positions. The index range \(-1 \leq i,j,\ell \leq N+2\) is adopted to accommodate boundary extrapolation.
To enforce the homogeneous Neumann boundary condition \cref{boundary-large} while preserving high-order accuracy, a third-order extrapolation scheme is implemented. For instance, along the \(z\)-direction (normal to the boundary at \(z=0\) and \(z=L\)), the extrapolation rules for the magnetization field are:
\begin{align*}
\m_{i,j,1}=\m_{i,j,0},\quad \m_{i,j,-1}=\m_{i,j,2},\quad \m_{i,j,N+1}=\m_{i,j,N} ,\quad \m_{i,j,N+2}=\m_{i,j,N-1}.
\end{align*}
Analogous extrapolation formulas are applied to enforce the boundary condition along the \(x\)- and \(y\)-directions, ensuring consistent high-order accuracy across the entire computational domain.

The fourth-order spatial difference operators should be introduced further. 
To achieve fourth-order spatial accuracy, which is essential for resolving fine magnetic structures (e.g., domain walls with nanoscale width) and minimizing discretization-induced errors, we employ long-stencil finite difference operators for first and second partial derivatives. For the \(x\)-direction, the fourth-order accurate operators for \(\partial_x\) (denoted \({\mathcal D}_{x,(4)}^1\)) and \(\partial_x^2\) (denoted \({\mathcal D}_{x,(4)}^2\)) are defined as:
\begin{eqnarray} 
\hspace{-0.35in}  
{\mathcal D}_{x,(4)}^1 f_{i,j,k} &=& \tilde{D}_x ( 1 - \frac{h_x^2}{6} D_x^2 ) f_{i,j,k} \nonumber 
\\
&=& 
\frac{  f_{i-2,j,k} - 8 f_{i-1,j,k}  + 8 f_{i+1,j,k} - f_{i+2,j,k} }{12 h_x} ,  
\label{FD-4th-1} 
\\
\hspace{-0.35in}  
{\mathcal D}_{x,(4)}^2 f_{i,j,k} &=& D_x^2 ( 1 - \frac{h_x^2}{12} D_x^2 ) f_{i,j,k}  \nonumber 
\\
&=& 
\frac{ - f_{i-2,j,k} + 16 f_{i-1,j,k,k} - 30 f_{i,j,k} + 16 f_{i+1,j,k} - f_{i+2,j,k} }{12 h_x^2 } . 
\label{FD-4th-2} 
\end{eqnarray} 
These operators are derived by eliminating leading-order discretization errors through the inclusion of neighboring grid points, resulting in fourth-order convergence (\(O(h^4)\)) for sufficiently smooth functions.

By symmetry, the fourth-order difference operators for the \(y\)- and \(z\)-directions—designated \({\mathcal D}_{y,(4)}^1\), \({\mathcal D}_{y,(4)}^2\), \({\mathcal D}_{z,(4)}^1\), and \({\mathcal D}_{z,(4)}^2\)—are defined by substituting the respective spatial coordinates and mesh-sizes. The discrete Laplacian operator \(\Delta_h\), which approximates the continuous Laplacian \(\Delta\) with fourth-order accuracy, is constructed as the sum of the second-order operators in all three directions: \(\Delta_h = {\mathcal D}_{x,(4)}^2 + {\mathcal D}_{y,(4)}^2 + {\mathcal D}_{z,(4)}^2\).

To establish a rigorous performance baseline for the proposed numerical method, we revise the first-order (BDF1) and second-order (BDF2) backward differentiation formula (BDF) schemes, which are canonical methods in micromagnetics, to incorporate the fourth-order spatial operators and the equivalent LLG formulation derived in \cref{eq-model}. These revised BDF schemes will serve as benchmark references in subsequent numerical experiments, enabling quantitative assessment of the proposed method's accuracy, efficiency, and stability.

\subsection{The first order method}

The BDF1 algorithm is given below:
\begin{equation}\label{BDF1}
\left\{ 
\begin{aligned}
&\frac{ {\tilde{\m}}_h^{n+1} -  {\m}_h^n}{k}
=-{\m}_h^{n} \times (\epsilon \Delta_{h}\tilde{\m}_h^{n+1}+{\f}_h^{n})+\alpha(\epsilon \Delta_{h}\tilde{\m}_h^{n+1} +{\f}_h^{n})\\
&\qquad+\alpha(\epsilon |{\nabla}_{h}{\m}_h^{n}|^2+{\m}_h^{n}\cdot {\f}_h^{n}) {\m}_h^{n}, \\
& \qquad\qquad\qquad\qquad\quad \m_h^{n+1} = \frac{\tilde{\m}_h^{n+1}}{ |\tilde{\m}_h^{n+1}| },
\end{aligned}
\right.
\end{equation}

\subsection{The second order method}
The BDF2 algorithm is formulated below:
\begin{equation}\label{sipm}
\left\{ 
\begin{aligned}
&\frac{\frac32 {\tilde{\m}}_h^{n+2} - 2 {\m}_h^{n+1} + \frac12 {\m}_h^n}{k}
=-\hat{\m}_h^{n+2} \times (\epsilon \Delta_{h}\tilde{\m}_h^{n+2}+\hat{\f}_h^{n+2})+\alpha(\epsilon \Delta_{h}\tilde{\m}_h^{n+2} +\hat{\f}_h^{n+2})\\
&\qquad+\alpha(\epsilon |\tilde{\nabla}_{h}\hat{\m}_h^{n+2}|^2+\hat{\m}_h^{n+2}\cdot \hat{\f}_h^{n+2}) \hat{\m}_h^{n+2}, \\
& \qquad\qquad\qquad\qquad\quad \m_h^{n+2} = \frac{\tilde{\m}_h^{n+2}}{ |\tilde{\m}_h^{n+2}| },
\end{aligned}
\right.
\end{equation} 
where $\tilde{\m}_h^{n+2}$ is an intermediate magnetization, and $\hat{\m}_h^{n+2}$, $\hat{\f}_h^{n+2}$ are given by the following extrapolation formula: 
\begin{align*}
\hat{\m}_h^{n+2} &=2{\m}_h^{n+1}-{\m}_h^n, \label{m_hat}\\
\hat{\f}_h^{n+2} &=2{\f}_h^{n+1}-{\f}_h^n,
\end{align*}
with $\f_h^{n}=-Q(m_2^n\e_2+m_3^n\e_3)+\h_s^n+\h_e^n$. The presence of cross product yields a linear system of equations with non-symmetric structure and variable coefficients. In turn, the GMRES solver has to be applied to implement this numerical system. The numerical evidence has revealed that, the convergence of GMRES solver becomes slower for larger temporal step-size $k$ or smaller spatial grid-size $h$, which makes the computation more challenging. 

\subsection{The proposed numerical method} \label{discretisations}

The BDF1 method (detailed in \cref{BDF1}) and BDF2 method (formulated in \eqref{sipm}) both employ a semi-implicit approach for handling the gyromagnetic and damping terms in micromagnetic simulations. Specifically, the Laplacian term \(\Delta \m\) is discretized implicitly, while coefficient functions are updated using a third-order accurate explicit extrapolation formula. Given that the magnetization vector \(\m\) maintains a constant magnitude of 1, the strength of the gyromagnetic term is governed by \(\Delta\m +\f\); in contrast, the damping term’s strength is determined by the product of \(\Delta\m +\f\) and the Gilbert damping parameter \(\alpha\).
A key limitation of these existing schemes is their constrained temporal accuracy: BDF1 achieves only first-order temporal accuracy, while BDF2 reaches second-order. This motivates a natural extension of the semi-implicit projection framework to develop a method with enhanced third-order temporal accuracy and fourth-order spatial accuracy, which forms the core objective of the proposed numerical scheme.
To streamline the presentation of the proposed method, we take the governing model \eqref{eq-model} as the starting point, and outline the numerical algorithm as follows.
\begin{equation}\label{proposed}
\left\{ 
\begin{aligned}
&\frac{\frac{11}{6}\tilde{\m}_h^{n+3}-3{\m}_h^{n+2}+\frac32{\m}_h^{n+1}-\frac13 {\m}_h^n}{k}\\
&\quad=-\hat{\m}_h^{n+3} \times (\epsilon \Delta_{h,(4)}\tilde{\m}_h^{n+3}+\hat{\f}_h^{n+3})+\alpha(\epsilon \Delta_{h,(4)}\tilde{\m}_h^{n+3} +\hat{\f}_h^{n+3})\\
&\qquad+\alpha(\epsilon |\tilde{\nabla}_{h,(4)}\hat{\m}_h^{n+3}|^2+\hat{\m}_h^{n+3}\cdot \hat{\f}_h^{n+3}) \hat{\m}_h^{n+3}, \\
&\m_h^{n+3}=\frac{\tilde{\m}_h^{n+3}}{|\tilde{\m}_h^{n+3}|},
\end{aligned}
\right.
\end{equation}
where
\begin{align*}
\hat{\m}_h^{n+3} &= 3 \m_h^{n+2}-3\m_h^{n+1} + \m_h^n,\\
\hat{\f}_h^{n+3} &= 3 \f_h^{n+2}-3\f_h^{n+1} + \f_h^n.
\end{align*}

\cref{tab-features} provides a comprehensive comparison of the proposed method, BDF2, and BDF1, focusing on key computational and structural metrics: number of unknowns, dimensional size of linear systems, symmetry patterns, availability of FFT-based fast solvers for linear equations, and frequency of stray field updates.
From a formal performance perspective, the proposed method exhibits clear superiority over both BDF1 and BDF2, particularly in terms of accuracy and computational efficiency. As elaborated in the numerical results presented in \cref{sec:experiments}, the proposed scheme proves to be a reliable and robust tool for micromagnetic simulations, delivering high accuracy and efficiency specifically in the regime of moderate damping parameters.
\begin{table}[htbp]
	\begin{center}
		\caption{Comparison of the proposed method, the BDF2 method, and the BDF1 method.}\label{tab-features}
		\begin{tabular}{cccc}
			\hline
			Property or number & Proposed method & BDF2 & BDF1\\
			\hline
			Linear systems& \boldsymbol{$1$} & $1$ & $1$ \\
			Size & \boldsymbol{$3N^3$} & 3$N^3$& $3N^3$ \\
			Symmetry& {\bf No}& No& No \\
			Fast Solver& {\bf No}& No& No \\
			Accuracy& \boldsymbol{$\mathcal{O}(k^3+h^4)$} & $\mathcal{O}(k^2+h^2)$ & $\mathcal{O}(k+h^2)$ \\
			Stray field updates & \boldsymbol{$1$} &$1$ &$1$ \\
			\hline
		\end{tabular}
	\end{center}
\end{table}

\section{Numerical experiments}
\label{sec:experiments}

This section presents numerical experiments to systematically evaluate three micromagnetic methods: the proposed third-order scheme, BDF2 (second-order), and BDF1 (first-order). To cover practical scenarios, experiments use a designed sequence of Gilbert damping parameters, spanning weak, moderate, and strong dissipation. Performance is assessed via three core metrics: accuracy, efficiency, and stability, with detailed data for direct comparison. Accuracy is verified using exact solutions and convergence rates; efficiency via wall-clock time and computational error; stability by monitoring magnetization profiles and energy dissipation. Special focus is given to domain wall dynamics (critical for magnetic devices). Domain wall velocity—an important parameter—is recorded and analyzed against damping and external magnetic field. This validates method consistency and reveals damping-field-velocity interplay, aiding magnetic device optimization.

\subsection{Numerical accuracy and efficiency}

For the purpose of simplifying the temporal accuracy analysis, we set the parameter \(\epsilon=1\) and the forcing term \(\f=0\) in the governing model \cref{eq-model}. To serve as the benchmark for error quantification, analytical exact solutions are derived for both one-dimensional (1D) and three-dimensional (3D) scenarios.

In the 1D case, the exact magnetization solution \(\m_e\) is expressed as:
\begin{equation*}
	\m_e=\left(\cos(\cos(\pi x))\sin t, \sin(\cos(\pi x))\sin t, \cos t\right)^T,
\end{equation*}
whereas the corresponding 3D exact solution takes the form:
\begin{equation*}
\m_e=\left(\cos(\cos(\pi x)\cos(\pi y)\cos(\pi z))\sin t, \sin(\cos(\pi x)\cos(\pi y)\cos(\pi z))\sin t, \cos t\right)^T,
\end{equation*}
It is verified that the above exact solutions satisfy the governing equation \cref{eq-model} when the forcing term is defined as \(\g=\partial_t \m_e-\alpha \Delta \m_e -\alpha |\nabla \m_e|^2+\m_e \times \Delta \m_e\). Additionally, these solutions adhere to the homogeneous Neumann boundary condition, ensuring consistency with the simulation constraints.

To isolate the temporal approximation error from spatial discretization effects, the spatial resolution in the 1D test is fixed at \(h=10^{-4}\)—a sufficiently fine grid that renders the spatial approximation error negligible compared to the temporal error. The Gilbert damping parameter is set to \(\alpha=0.01\), and the simulation is run until the final time \(T=0.1\). Under this configuration, the error quantified primarily reflects the inaccuracy introduced by temporal discretization.

The 3D temporal accuracy test faces inherent constraints from spatial resolution, as excessively fine spatial grids would lead to prohibitive computational costs. To balance spatial and temporal error contributions, we adopt a coordinated refinement strategy for spatial mesh sizes (\(h_x, h_y, h_z\)) and temporal step-size (\(k\)) tailored to the order of each numerical method:
\begin{itemize}
	\item - For the first-order BDF1 method: \(k=h_x^2=h_y^2=h_z^2=h^2=T/N_0\);
	\item - For the second-order BDF2 method: \(k=h_x=h_y=h_z=h=T/N_0\);
	\item - For the proposed third-order method: \(k=h_x^{4/3}=h_y^{4/3}=h_z^{4/3}=h^{4/3}=T/N_0\).
\end{itemize}
Here, \(N_0\) denotes the refinement level parameter, whose specific values are detailed in subsequent result presentations. Consistent with the 1D test, the damping parameter is set to \(\alpha=0.01\), while the final simulation time \(T\) is specified in the corresponding result section.

The numerical errors of the three methods (BDF1, BDF2, and the proposed method) are systematically recorded as a function of the temporal step-size \(k\), and the detailed data are compiled in \cref{tab-1}. From the error analysis, the temporal accuracy orders of the methods are clearly identified: the proposed method achieves third-order (\(3\)) temporal accuracy, while BDF1 and BDF2 exhibit first-order (\(1\)) and second-order (\(2\)) accuracy, respectively. Notably, this order consistency holds true for both 1D and 3D computational scenarios, confirming the reliability of the proposed method across different spatial dimensions.

\begin{table}[htbp] 
	\centering
	{\caption{The numerical errors for the proposed method, the BDF1 and the BDF2 with $\alpha=0.01$ and $T=0.1$. Left: 1D with $h=1D-4$; Right: 3D with $k=h_x^2=h_y^2=h_z^2=h^2=T/N_0$ for BDF1 and $k=h_x=h_y=h_z=h=T/N_0$ for the BDF2 method, and $k=h_x^{4/3}=h_y^{4/3}=h_z^{4/3}=h^{4/3}=T/N_0$ for the proposed method, with $N_0$ specified in the table.}\label{tab-1} }{
		\subfloat[Proposed method]{\label{tab:floatrow:one}%
			\begin{tabular}{cccc|cccc} 
				\hline	
				1D  & & {} & {} &3D &{} & {} &{} \\
				$k$ & $\|\cdot\|_{\infty}$ & $\|\cdot\|_{2}$ & $\|\cdot\|_{H^1}$ & 	$k,k^3\approx h^4$ & $\|\cdot\|_{\infty}$ & $\|\cdot\|_{2}$ & $\|\cdot\|_{H^1}$ \\
				\hline
			$T/12$ & 1.154D-9 & 9.135D-10 & 3.285D-9  & $T/6$ & 3.111D-6 & 5.690D-7 & 1.000D-5 \\
			$T/16$ & 5.075D-10 & 3.963D-10& 1.509D-9& $T/7$ & 1.950D-6 & 3.889D-7 & 7.230D-6\\
			$T/24$ & 1.553D-10 & 1.205D-10 & 4.840D-10 &$T/8$ & 1.043D-6 & 2.362D-7 & 4.579D-6\\
			$T/32$ & 6.867D-11 & 5.195D-11 & 2.161D-10  & $T/9$& 7.359D-7 & 1.748D-7 & 3.458D-6\\
			$T/36$ & 4.744D-11 & 3.633D-11 & 1.610D-10 & $T/11$& 4.430D-7 & 9.233D-8 & 1.828D-6\\
				order & 2.90& 2.93&2.76& {--}& 3.30&3.03 &2.83 \\
				
				
				%
				\hline
			\end{tabular}	
		}
		\qquad
		\subfloat[BDF1]{\label{tab:floatrow:two}
			\begin{tabular}{cccc|cccc} 
				\hline
				1D & &  & {} & 3D & {} &  & {} \\
				$k$ & $\|\cdot\|_{\infty}$ & $\|\cdot\|_{2}$ & $\|\cdot\|_{H^1}$ & $k=h^2$ & $\|\cdot\|_{\infty}$ & $\|\cdot\|_{2}$ & $\|\cdot\|_{H^1}$\\
				\hline
					$T/8$& 2.572D-5 & 2.341D-5 & 7.982D-5 & $T/40$ & 4.147D-4 & 6.722D-5& 5.380D-4 \\	
				$T/12$ & 1.775D-5 & 1.516D-5 & 5.584D-5 & $T/57$ & 2.900D-4 & 4.682D-5 & 3.729D-4 \\
				$T/16$ & 1.358D-5 & 1.125D-5 & 4.320D-5& $T/78$ & 2.139D-4 & 3.446D-5 & 2.736D-4\\
				$T/24$ & 9.239D-6 & 7.443D-6 & 2.980D-5 &$T/102$ & 1.642D-4 & 2.642D-5 & 2.093D-4\\
				$T/32$ & 7.003D-6 & 5.565D-6 & 2.276D-5  & $T/129$& 1.300D-4 & 2.089D-5 & 1.652D-4\\
				order &0.94 &1.03&0.91 &{--}& 0.99&1.00&1.01 \\
				
				
				%
				\hline
			\end{tabular}
		}
		\qquad
		\subfloat[BDF2]{\label{tab:floatrow:three}
			\begin{tabular}{cccc|cccc} 
				\hline
				1D  & & & {} & 3D & {} &  & {}\\
				$k$ & $\|\cdot\|_{\infty}$ & $\|\cdot\|_{2}$ & $\|\cdot\|_{H^1}$ & $k=h$ & $\|\cdot\|_{\infty}$ & $\|\cdot\|_{2}$ & $\|\cdot\|_{H^1}$\\
				\hline
					$T/8$& 7.965D-6 & 6.676D-6 & 3.143D-5 & $T/2$ & 2.255D-4 & 3.235D-5  & 7.516D-4 \\	
				$T/12$ & 3.657D-6 & 3.049D-6 & 1.410D-5 & $T/3$ & 1.132D-4 & 1.884D-5 & 3.244D-4 \\
				$T/16$ & 2.083D-6 & 1.735D-6 & 7.953D-6& $T/4$ &  6.622D-5 & 1.175D-5 & 1.815D-4\\ 
				$T/24$ & 9.355D-7 & 7.789D-7 & 3.541D-6 &$T/5$ & 4.312D-5 & 7.923D-6  & 1.163D-4\\
				$T/32$ & 5.283D-7 & 4.401D-7 & 1.993D-6 & $T/6$& 3.058D-5 & 5.680D-6 & 8.090D-5\\
				order & 1.96&1.96&1.99 &{--}&1.82 &1.59& 2.03\\
				
				
				%
				\hline
			\end{tabular}
	} }	
\end{table}

Following the temporal accuracy evaluation, spatial accuracy tests were conducted to quantify the spatial discretization performance of the proposed third-order method, alongside the BDF1 and BDF2 schemes. To ensure that temporal approximation errors do not interfere with the spatial accuracy assessment, the temporal step-size was fixed at a sufficiently small value of \(k=10^{-5}\) across all 1D and 3D tests—this choice renders temporal errors negligible compared to the spatial errors being measured.
Consistent with the temporal accuracy test configuration, the Gilbert damping parameter was set to \(\alpha=0.01\), and all simulations were executed up to a final time of \(T=0.1\). The exact solutions detailed in the preceding section for 1D and 3D, respectively) were again employed as the benchmark to calculate numerical errors. Specifically, the \(L^2\)-norm error between the numerical solution and the exact solution was computed for a sequence of decreasing spatial grid-sizes \(h\), isolating the impact of spatial discretization.
The spatial error data for all three methods, organized by spatial grid-size \(h\), are systematically documented in \cref{tab-2}. A comprehensive analysis of these results reveals distinct spatial accuracy characteristics among the schemes: the BDF1 and BDF2 algorithms both achieve second-order spatial accuracy, while the proposed method demonstrates a higher fourth-order spatial accuracy. Importantly, this disparity in spatial accuracy orders is consistent across both 1D and 3D computational scenarios, further validating the superior spatial discretization capability of the proposed numerical method.

\begin{table}[htbp]
	\centering
	{\caption{The numerical errors of the proposed method, the BDF1 and the BDF2 with $\alpha=0.01$ and $T=0.1$. Left: 1D with $k=1D-5$; Right: 3D with $k=1D-5$.} \label{tab-2} }{
		\subfloat[Proposed method]{\label{tab:floatrow:1-S}
			\begin{tabular}{cccc|cccc}	
				\hline
				1D  & & & {} & 3D & & & \\
				$h$ & $\|\cdot\|_{\infty}$ &$\|\cdot\|_{2}$ &$\|\cdot\|_{H^1}$& $h$ & $\|\cdot\|_{\infty}$ & $\|\cdot\|_{2}$ & $\|\cdot\|_{H^1}$  \\
				\hline
				1/16 & 9.094D-6& 6.487D-6 & 9.920D-5& 1/4 & 1.466D-3& 3.971D-4 & 6.935D-3\\
				1/32 & 5.811D-7 & 4.117D-7 & 6.394D-6 & 1/6 & 3.509D-4 & 8.772D-5 & 1.574D-3\\
				1/64 & 3.647D-8 & 2.583D-8 & 4.028D-7 & 1/8 & 1.267D-4 & 2.897D-5 & 5.305D-4\\
				1/128 & 2.283D-9 & 1.616D-9 & 2.523D-8 & 1/10 & 5.825D-5 & 1.213D-5 & 2.244D-4 \\
				1/256 &  1.428D-10 & 1.010D-10 & 1.578D-9  & 1/12 & 2.987D-5 & 5.928D-6 & 1.101D-4 \\
				order  &3.99 &3.99&3.99 & {--} &3.54&3.83&3.77 \\
				\hline
			\end{tabular}
		}	
		\qquad
		\subfloat[BDF1]{\label{tab:floatrow:2-S}
			\begin{tabular}{cccc|cccc}	
				\hline
				1D  & &  & {} & 3D & & & \\
				$h$ & $\|\cdot\|_{\infty}$ &$\|\cdot\|_{2}$ &$\|\cdot\|_{H^1}$ & $h$ & $\|\cdot\|_{\infty}$ &$\|\cdot\|_{2}$ &$\|\cdot\|_{H^1}$\\
				\hline
				1/16 & 4.223D-4 & 2.893D-4 & 2.216D-3& 1/4 & 8.026D-3& 1.816D-3 & 1.589D-2\\
				1/32 & 1.060D-4 & 7.208D-5 & 5.503D-4 & 1/6 & 4.094D-3 & 7.779D-4 & 6.575D-3\\
				1/64 & 2.654D-5 & 1.800D-5 & 1.373D-4 & 1/8 & 2.437D-3 & 4.326D-4 & 3.570D-3\\
				1/128 & 6.636D-6 & 4.499D-6 & 3.432D-5 & 1/10 & 1.611D-3 & 2.753D-4 & 2.242D-3 \\
				1/256 &  1.658D-6 & 1.124D-6 & 8.580D-6  & 1/12 &1.139D-3 & 1.906D-4 & 1.539D-3 \\
				order  & 2.00&2.00&2.00 & {--} &1.78&2.05&2.13 \\
				\hline
			\end{tabular}
		}
		\qquad
		\subfloat[BDF2]{\label{tab:floatrow:3-S}
			\begin{tabular}{cccc|cccc}	
				\hline
				1D  & &  & & 3D & & & \\
				$h$ & $\|\cdot\|_{\infty}$ &$\|\cdot\|_{2}$ &$\|\cdot\|_{H^1}$ &$h$ & $\|\cdot\|_{\infty}$ &$\|\cdot\|_{2}$ &$\|\cdot\|_{H^1}$ \\
				\hline
				1/16 & 4.223D-4 & 2.894D-4 & 2.216D-3& 1/4 &8.027D-3 & 1.817D-3 & 1.589D-2 \\
				1/32 & 1.060D-4 &7.208D-5& 5.503D-4 & 1/6 & 4.094D-3 & 7.779D-4 & 6.576D-3 \\
				1/64 & 2.654D-5& 1.800D-5 & 1.373D-4 & 1/8 & 2.437D-3 & 4.326D-4 & 3.570D-3\\
				1/128 & 6.637D-6 & 4.500D-6 & 3.432D-5 & 1/10 & 1.611D-3 & 2.753D-4 & 2.242D-3 \\
				1/256 &  1.659D-6 & 1.125D-6 & 8.579D-6  & 1/12 & 1.139D-3 & 1.906D-4 & 1.539D-3 \\
				order  &2.00&2.00&2.00 & {--} &1.78 &2.05&2.13\\
				\hline
			\end{tabular} 
		}
	}
\end{table}

\begin{figure}[htbp]
	\centering
	\subfloat[Varying $k$ in 1D up to $T=0.1$ ]{\label{cputime_1D}\includegraphics[width=2.5in]{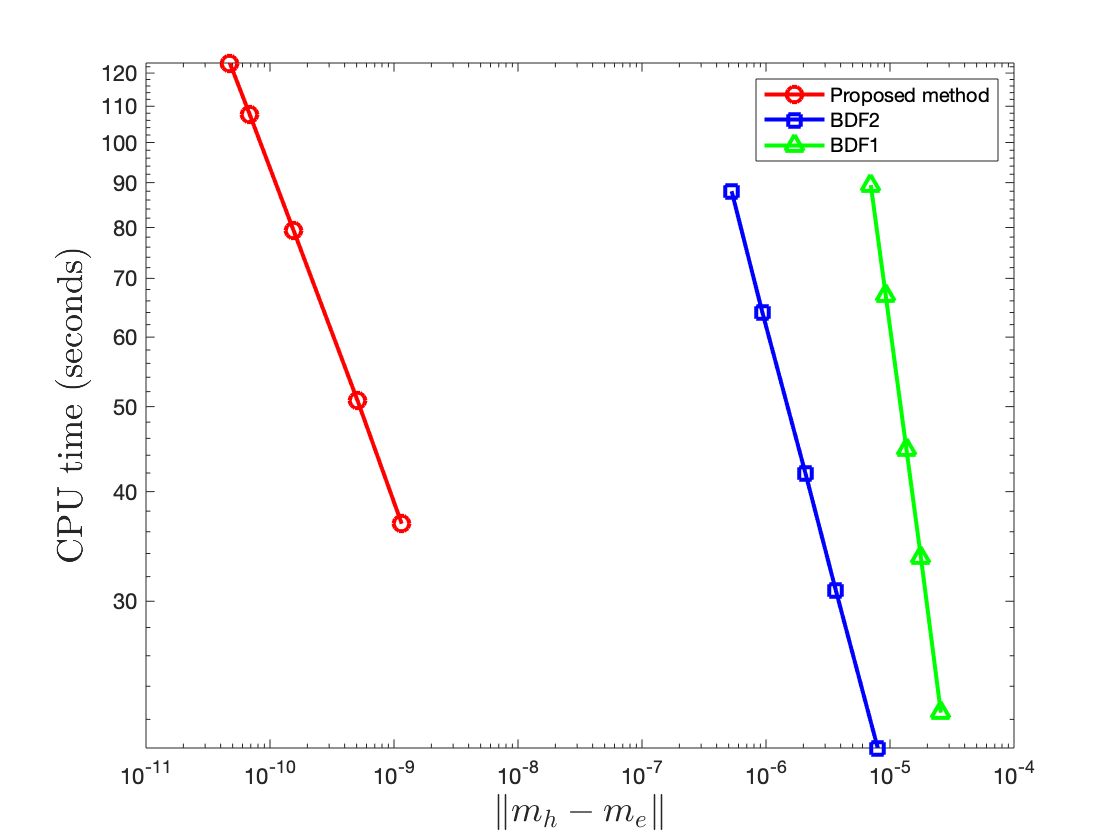}}
	\subfloat[Varying $k$ in 3D up to $T=0.1$]{\label{cputime_3D}\includegraphics[width=2.5in]{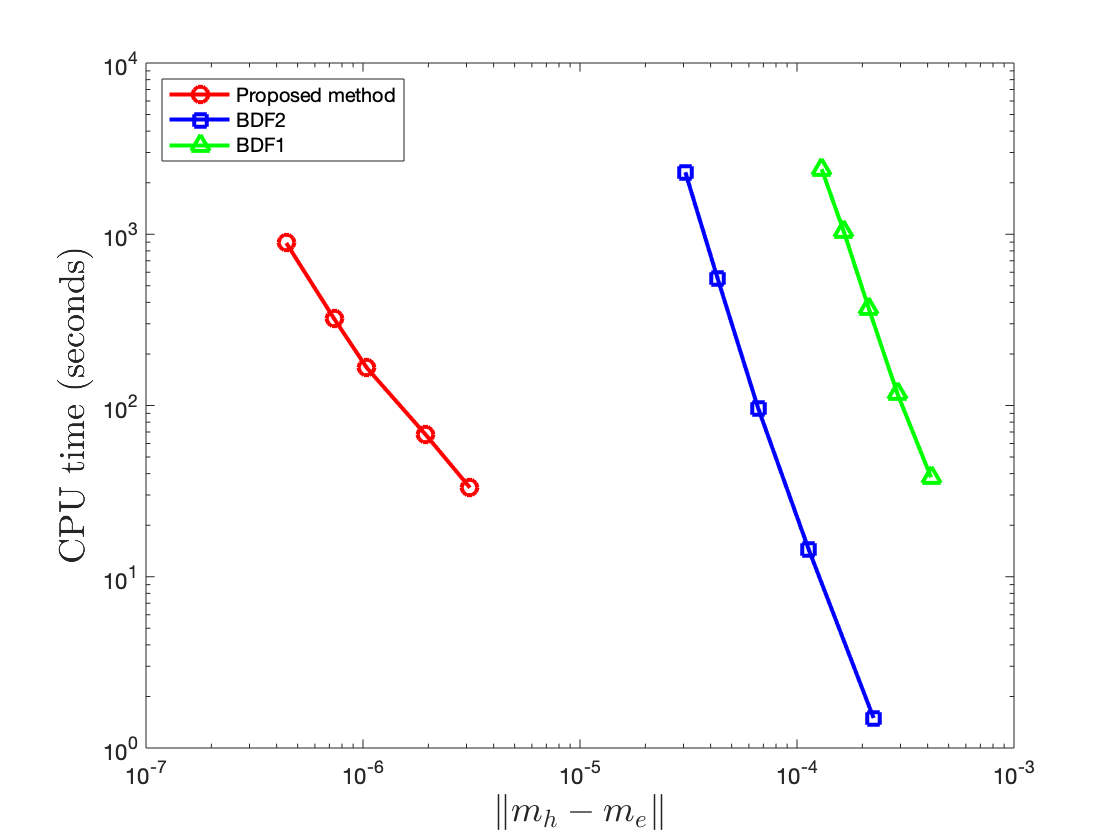}}
	\hspace{0.1in}
	\subfloat[Varying $h$ in 1D up to $T=0.1$]{\label{cputime_1D_space}\includegraphics[width=2.5in]{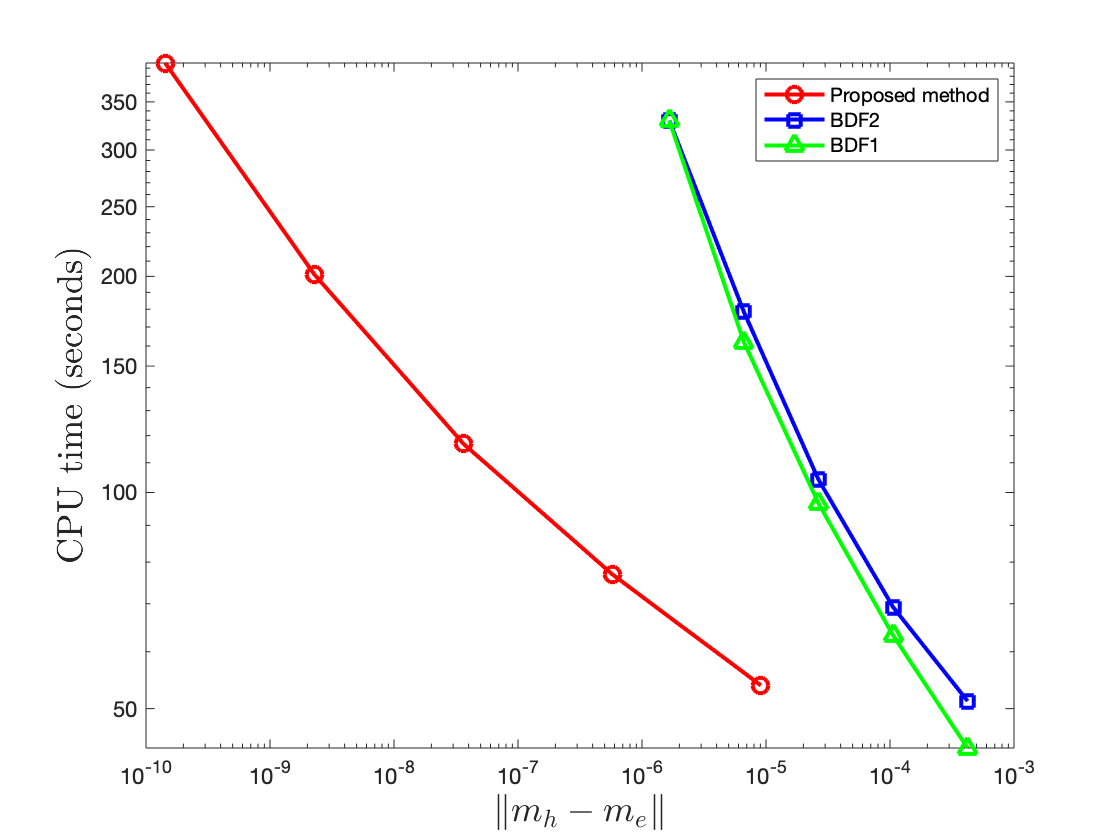}}
	\subfloat[Varying $h$ in 3D up to $T=0.1$]{\label{cputime_3D_space}\includegraphics[width=2.5in]{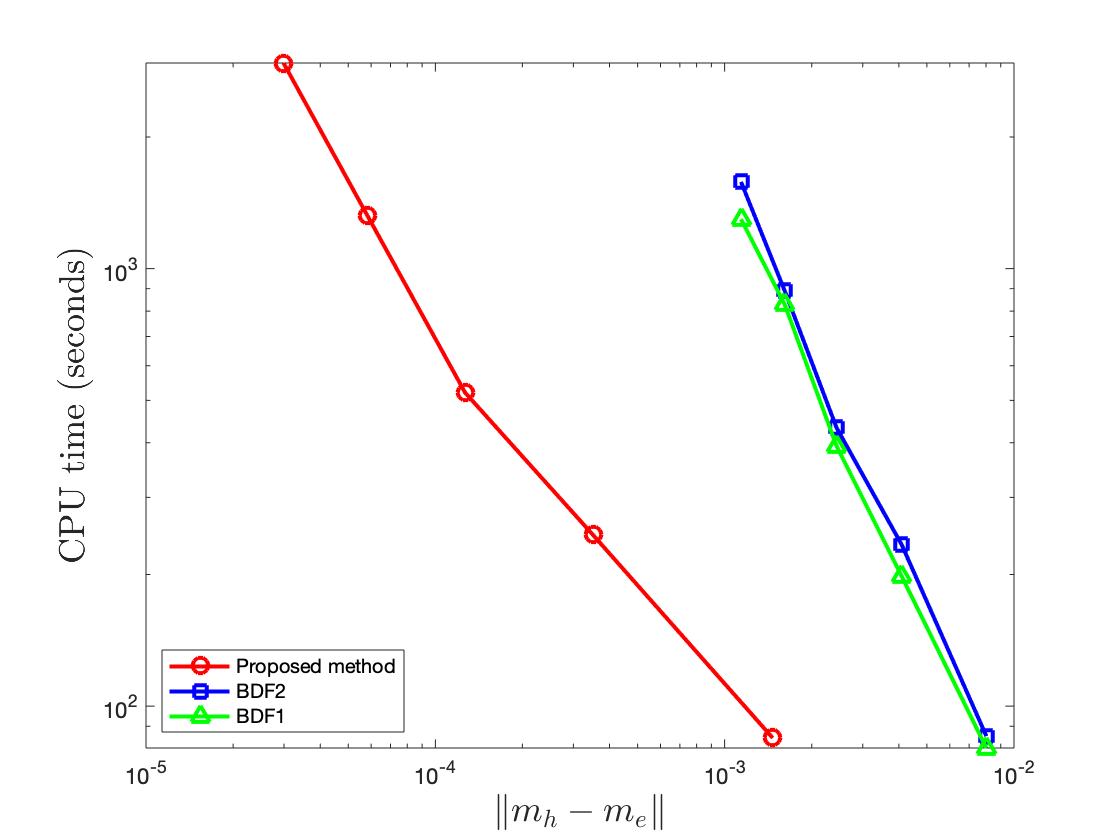}}
	\caption{CPU time needed to achieve the desired numerical accuracy, for the proposed method, the BDF2 and the BDF1 method, in both the 1D and 3D computations. The CPU time is recorded as a function of the approximation error by varying $k$ or $h$ independently. CPU time with varying $k$: proposed method $<$ BDF2 $<$ BDF1; CPU time with varying $h$: proposed method $<$ BDF1 $\lessapprox$ BDF2.}\label{cputime}
\end{figure}

To make a comparison in terms of the numerical efficiency, we plot the CPU time (in seconds) vs. the error norm $\|\m_h-\m_e\|_{\infty}$. In details, the CPU time is recorded as a function of the approximation error in \cref{cputime_1D} in 1D and in \cref{cputime_3D} in 3D, with a variation of $k$ and a fixed value of $h$. Similar plots are also displayed in \cref{cputime_1D_space} in 1D and \cref{cputime_3D_space} in 3D, with a variation of $h$ and a fixed value of $k$. In the case of a fixed spatial resolution $h$, the proposed method is significantly more efficient than the BDF1 and the BDF2 in both the 1D and 3D computations. The BDF2 is slightly more efficient than the BDF1, while such an advantage may vary for different values of $k$ and $h$. In the case of a fixed time step size $k$, the proposed method is slightly more efficient than the BDF2 and BDF1, in both the 1D and 3D computations, and the the cost of BDF1 is comparable to
that of the BDF2.

\subsection{Stability test with large damping parameters}
\begin{figure}[htbp]
	\centering
	\subfloat{\label{BDF3_alpha_0_ang}\includegraphics[width=1.3in]{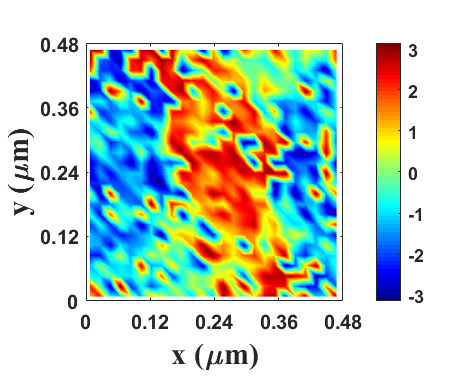}}
	\subfloat{\label{BDF3_alpha_0dot01_ang}\includegraphics[width=1.3in]{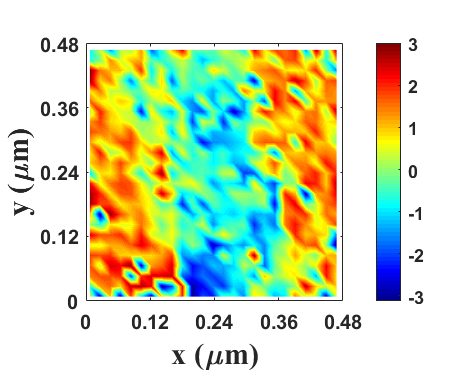}}
	\subfloat{\label{BDF3_alpha_0dot1_ang}\includegraphics[width=1.3in]{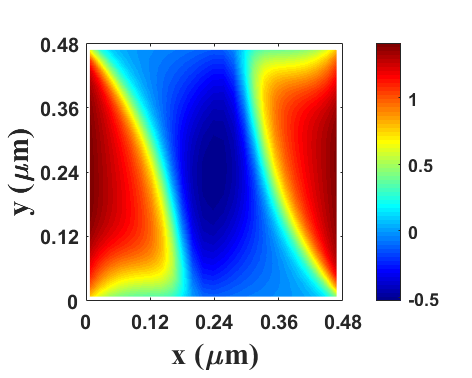}}
	\subfloat{\label{BDF3_alpha_1_ang}\includegraphics[width=1.3in]{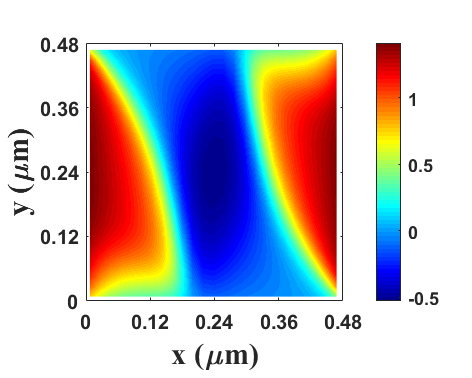}}
	\hspace{0.1in}
	\subfloat{\label{BDF3_alpha_5_ang}\includegraphics[width=1.3in]{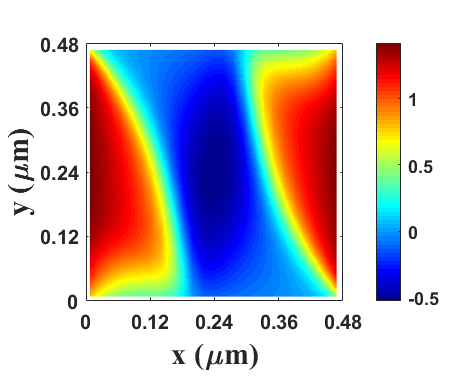}}
	\subfloat{\label{BDF3_alpha_10_ang}\includegraphics[width=1.3in]{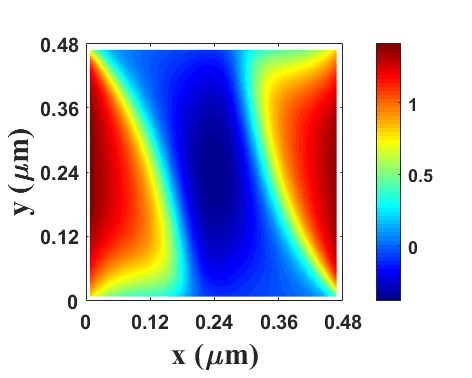}}
	\subfloat{\label{BDF3_alpha_40_ang}\includegraphics[width=1.3in]{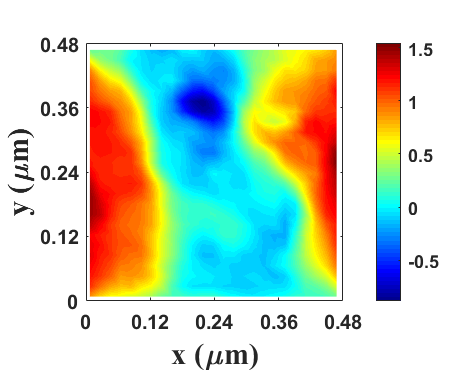}}
	\subfloat{\label{BDF3_alpha_100_ang}\includegraphics[width=1.3in]{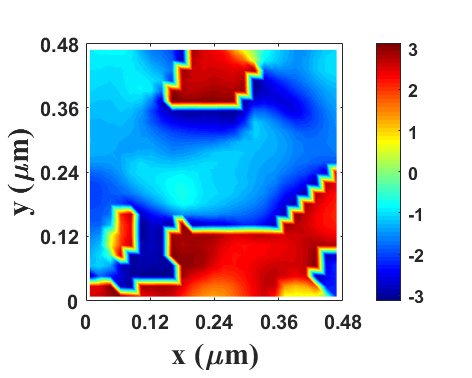}}
	\hspace{0.1in}
		\subfloat{\label{BDF2_alpha_0_ang}\includegraphics[width=1.3in]{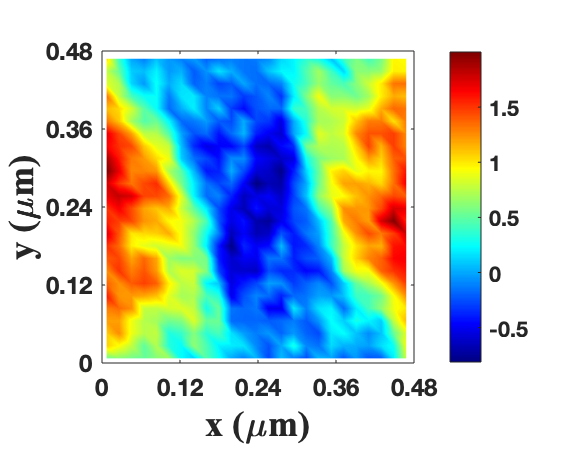}}
	\subfloat{\label{BDF2_alpha_0dot01_ang}\includegraphics[width=1.3in]{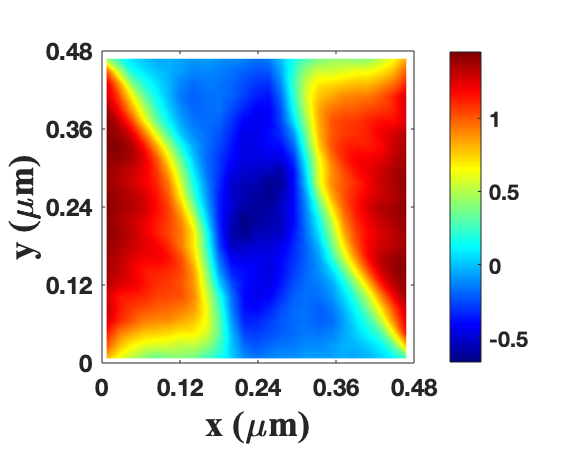}}
	\subfloat{\label{BDF2_alpha_0dot1_ang}\includegraphics[width=1.3in]{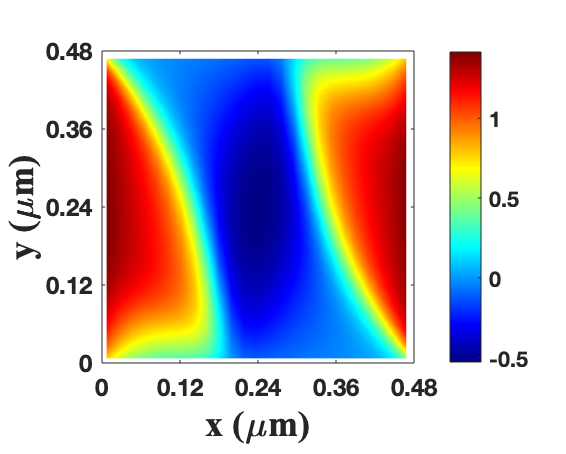}}
	\subfloat{\label{BDF2_alpha_1_ang}\includegraphics[width=1.3in]{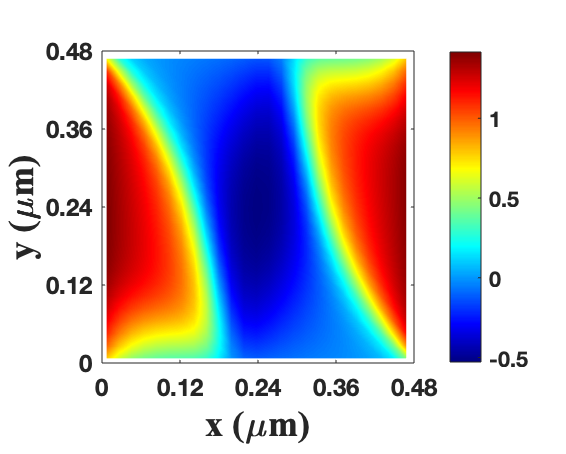}}
		\hspace{0.1in}
	\subfloat{\label{BDF2_alpha_5_ang}\includegraphics[width=1.3in]{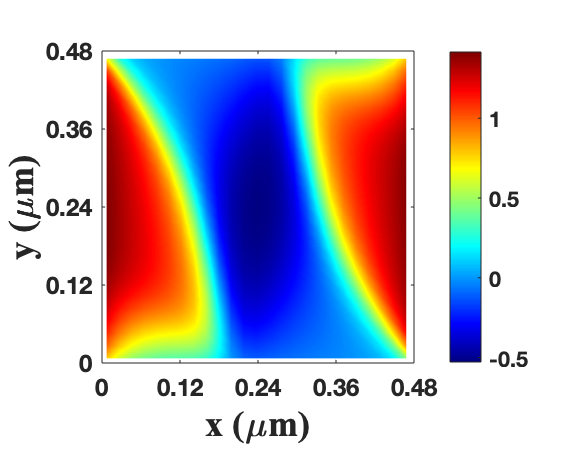}}
	\subfloat{\label{BDF2_alpha_10_ang}\includegraphics[width=1.3in]{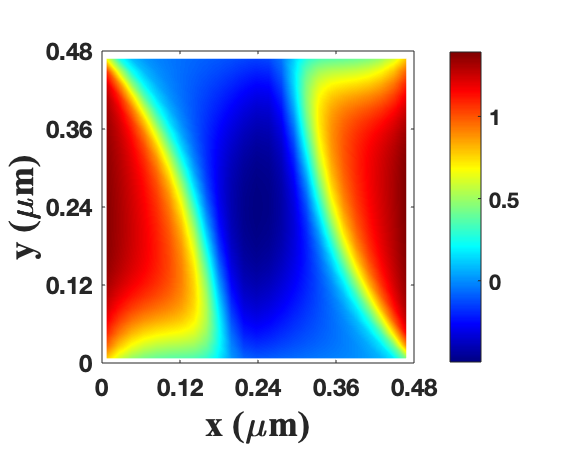}}
	\subfloat{\label{BDF2_alpha_40_ang}\includegraphics[width=1.3in]{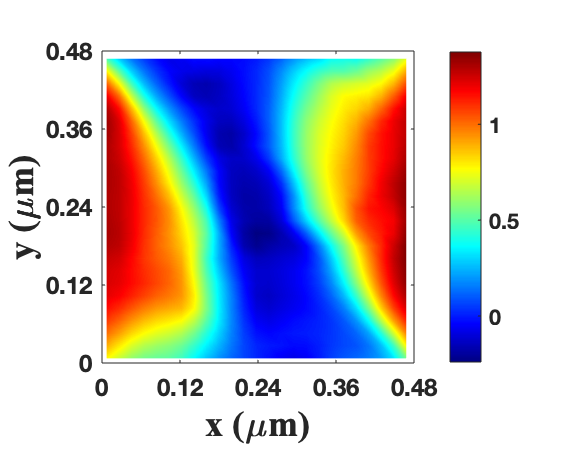}}
	\subfloat{\label{BDF2_alpha_100_ang}\includegraphics[width=1.3in]{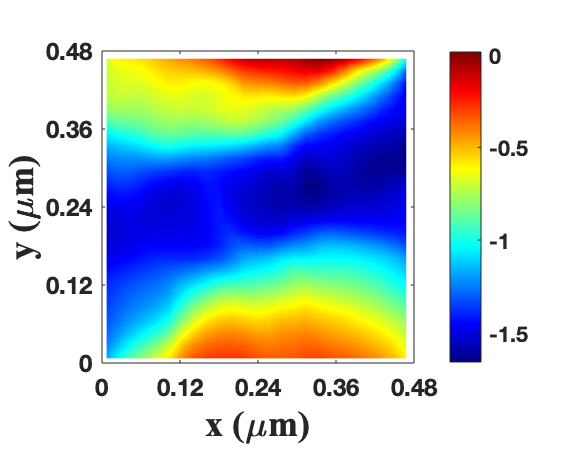}}
	\hspace{0.1in}
	\subfloat{\label{BDF1_alpha_0_ang}\includegraphics[width=1.3in]{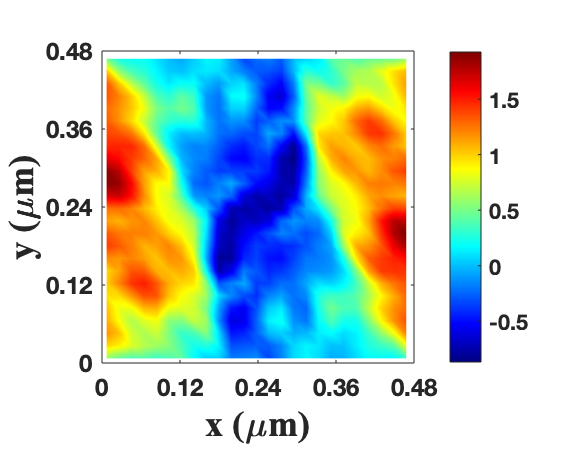}}
	\subfloat{\label{BDF1_alpha_0dot01_ang}\includegraphics[width=1.3in]{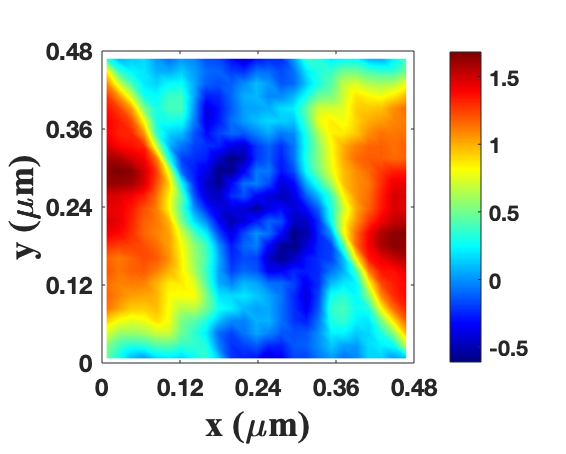}}
	\subfloat{\label{BDF1_alpha_0dot1_ang}\includegraphics[width=1.3in]{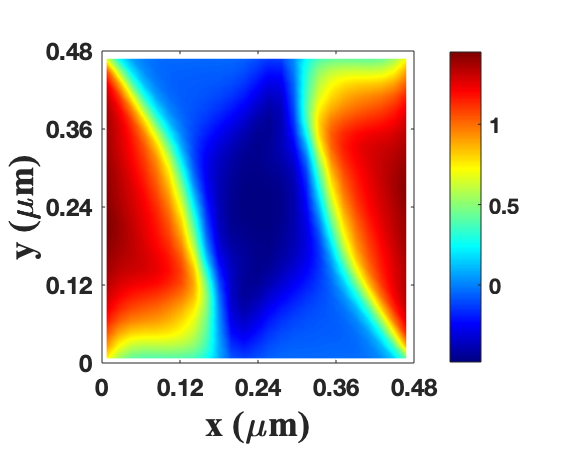}}
	\subfloat{\label{BDF1_alpha_1_ang}\includegraphics[width=1.3in]{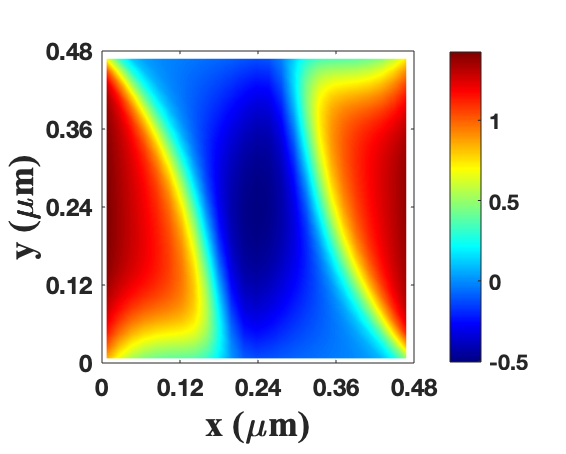}}
		\hspace{0.1in}
	\subfloat{\label{BDF1_alpha_5_ang}\includegraphics[width=1.3in]{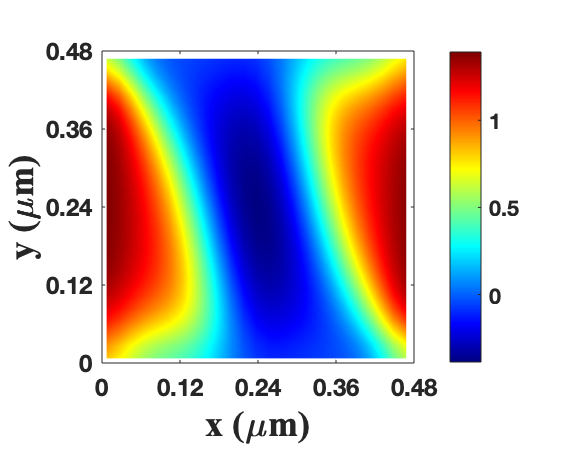}}
	\subfloat{\label{BDF1_alpha_10_ang}\includegraphics[width=1.3in]{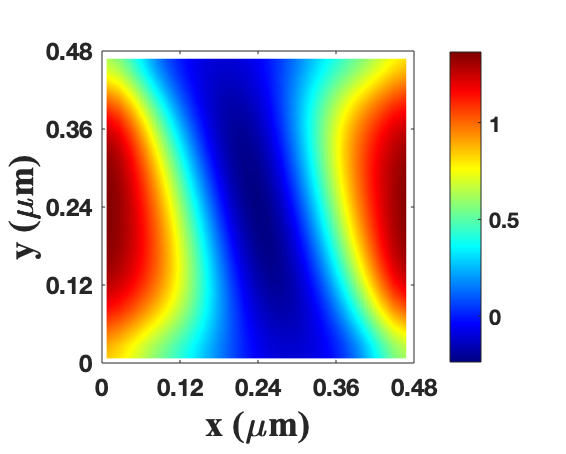}}
	\subfloat{\label{BDF1_alpha_40_ang}\includegraphics[width=1.3in]{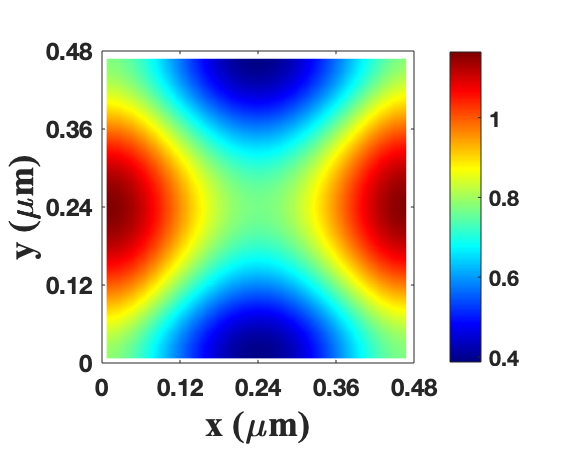}}
	\subfloat{\label{BDF1_alpha_100_ang}\includegraphics[width=1.3in]{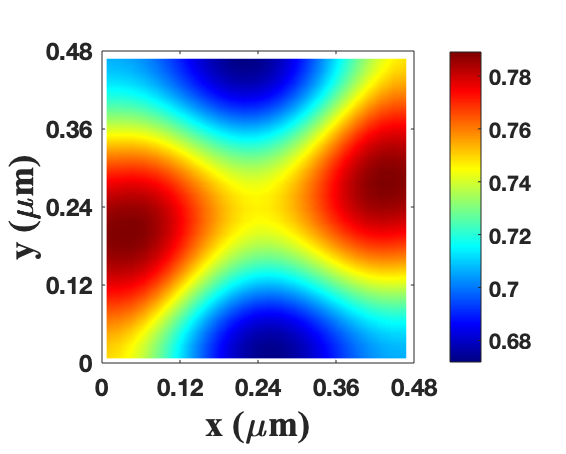}}
	\caption{Stable structures in the absence of magnetic field at $2\,$ns. The color denotes the angle between the first two components of the magnetization vector. Top two rows: Proposed method; Middle two rows: BDF2; Bottom two rows: BDF1. From left to right: $\alpha=0,0.01,0.1,1,5,10,40,100$. $dt=1\;ps$. }\label{BDF2_GSPM_alpha}
\end{figure}

\begin{figure}[htbp]
	\centering
	\subfloat{\label{BDF3_alpha_0_ang_v1}\includegraphics[width=1.3in]{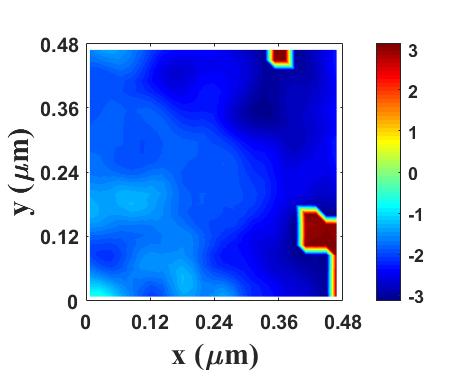}}
	\subfloat{\label{BDF3_alpha_0dot01_ang_v1}\includegraphics[width=1.3in]{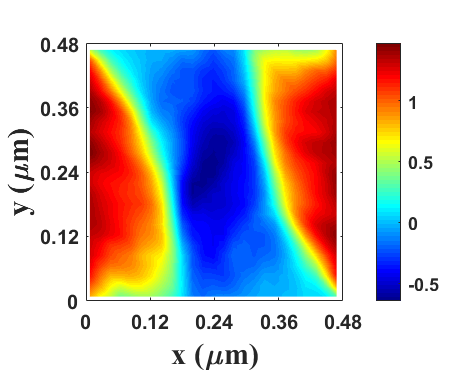}}
	\subfloat{\label{BDF3_alpha_0dot1_ang_v1}\includegraphics[width=1.3in]{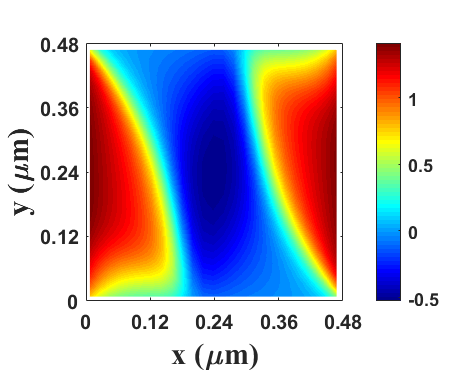}}
	\subfloat{\label{BDF3_alpha_1_ang_v1}\includegraphics[width=1.3in]{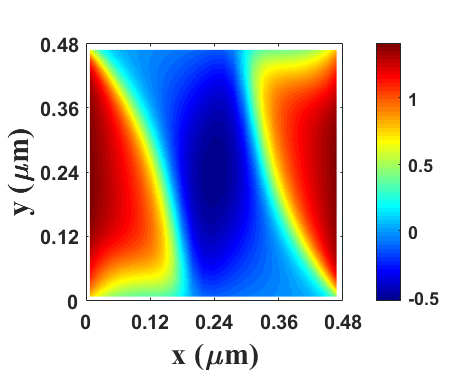}}
	\hspace{0.1in}
	\subfloat{\label{BDF3_alpha_5_ang_v1}\includegraphics[width=1.3in]{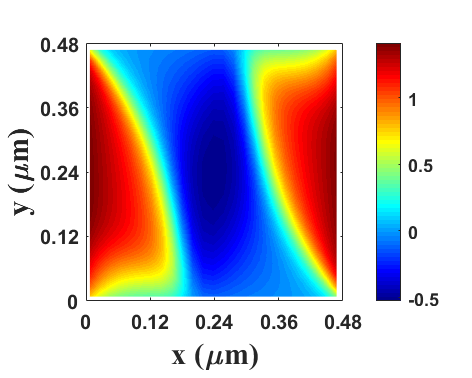}}
	\subfloat{\label{BDF3_alpha_10_ang_v1}\includegraphics[width=1.3in]{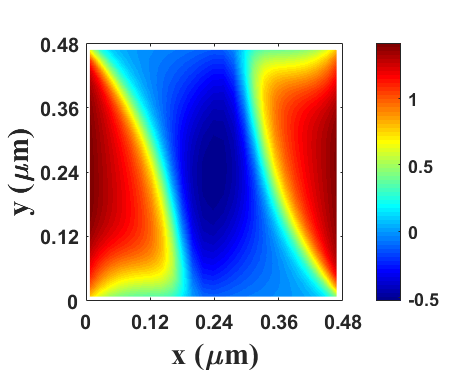}}
	\subfloat{\label{BDF3_alpha_40_ang_v1}\includegraphics[width=1.3in]{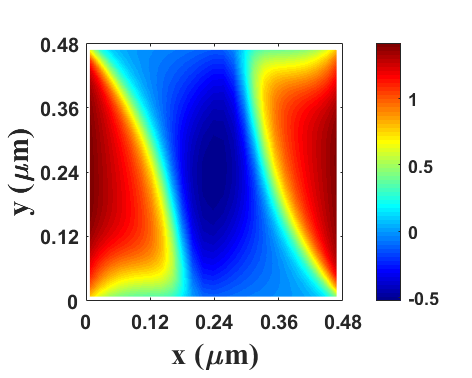}}
	\subfloat{\label{BDF3_alpha_100_ang_v1}\includegraphics[width=1.3in]{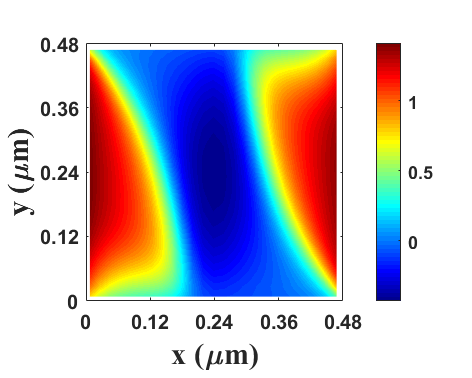}}
	\hspace{0.1in}
	\subfloat{\label{BDF2_alpha_0_ang_v1}\includegraphics[width=1.3in]{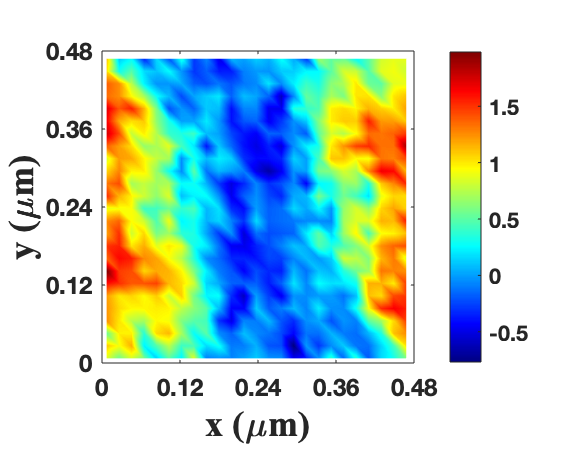}}
	\subfloat{\label{BDF2_alpha_0dot01_ang_v1}\includegraphics[width=1.3in]{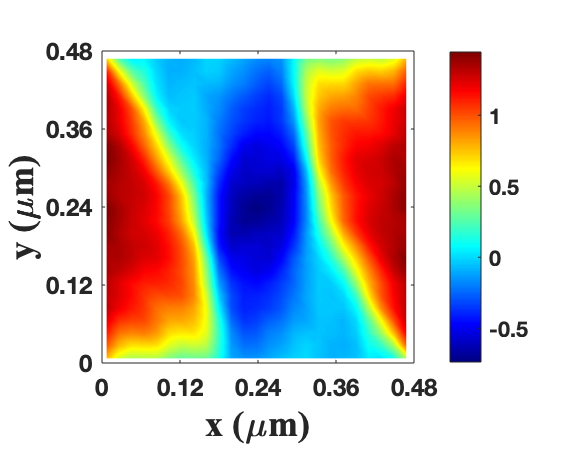}}
	\subfloat{\label{BDF2_alpha_0dot1_ang_v1}\includegraphics[width=1.3in]{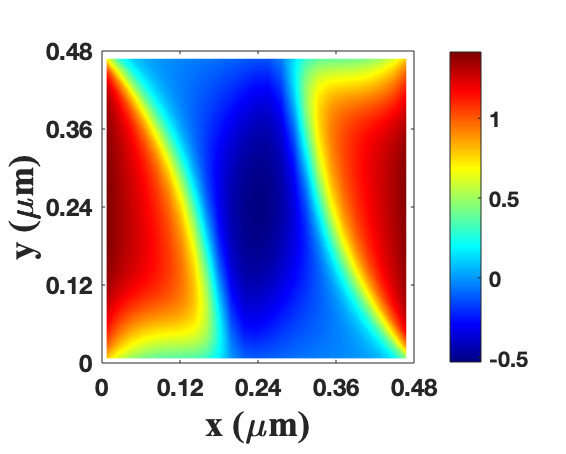}}
	\subfloat{\label{BDF2_alpha_1_ang_v1}\includegraphics[width=1.3in]{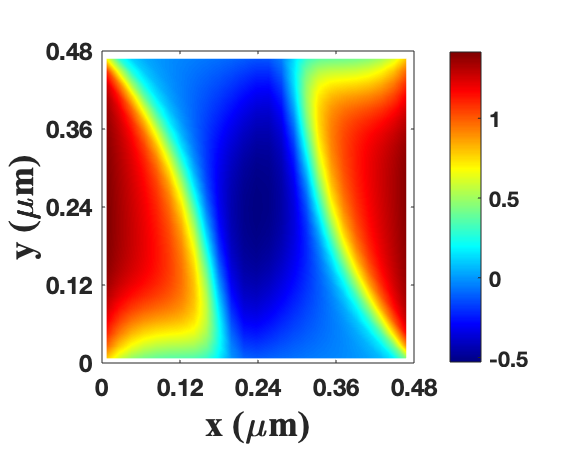}}
		\hspace{0.1in}
	\subfloat{\label{BDF2_alpha_5_ang_v1}\includegraphics[width=1.3in]{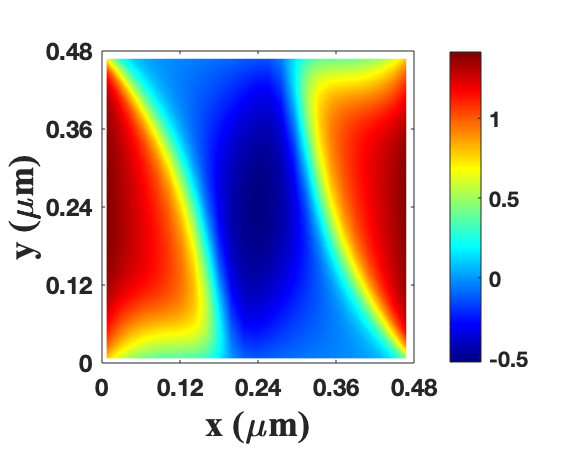}}
	\subfloat{\label{BDF2_alpha_10_ang_v1}\includegraphics[width=1.3in]{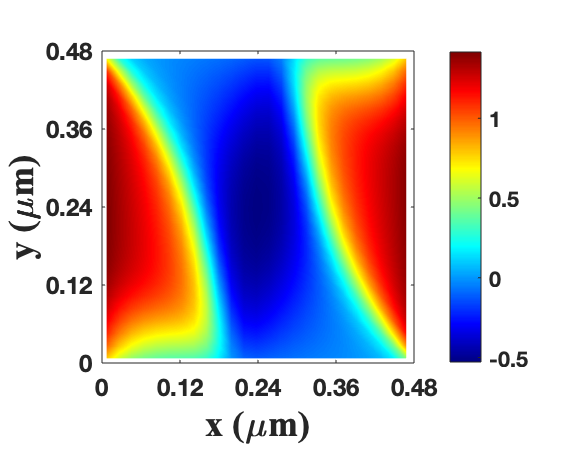}}
	\subfloat{\label{BDF2_alpha_40_ang_v1}\includegraphics[width=1.3in]{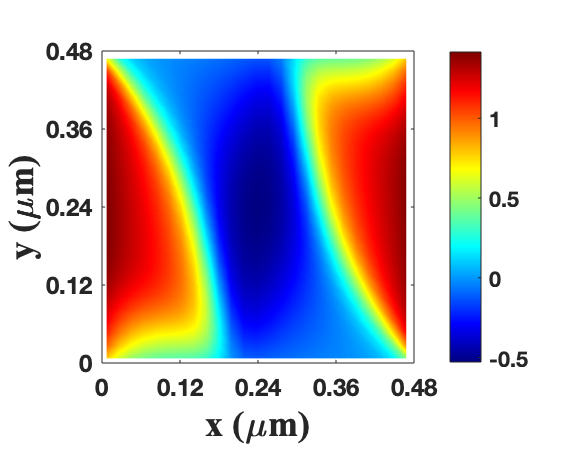}}
	\subfloat{\label{BDF2_alpha_100_ang_v1}\includegraphics[width=1.3in]{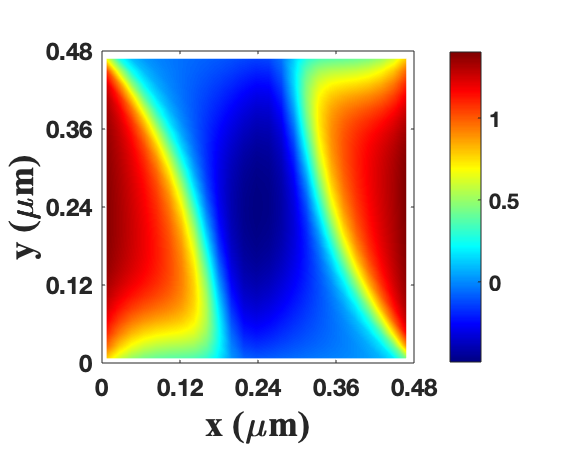}}
		\hspace{0.1in}
		\subfloat{\label{BDF1_alpha_0_ang_v1}\includegraphics[width=1.3in]{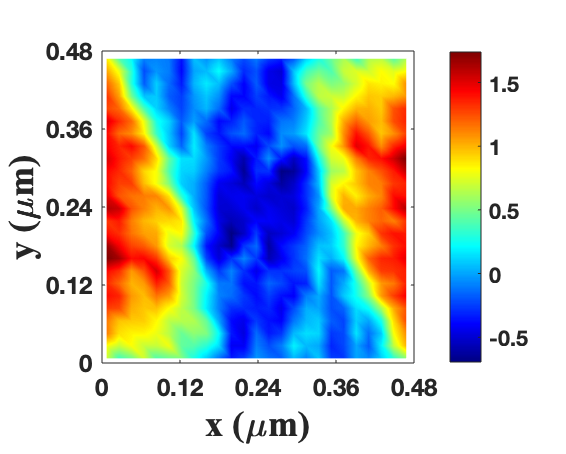}}
			\subfloat{\label{BDF1_alpha_0dot01_ang_v1}\includegraphics[width=1.3in]{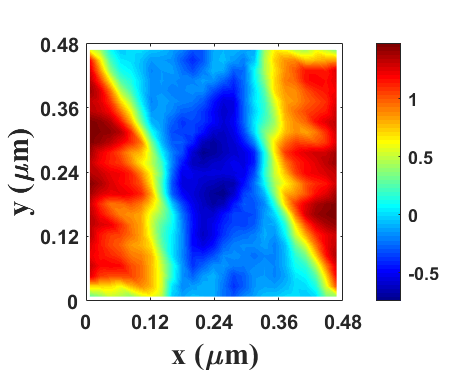}}
	\subfloat{\label{BDF1_alpha_0dot1_ang_v1}\includegraphics[width=1.3in]{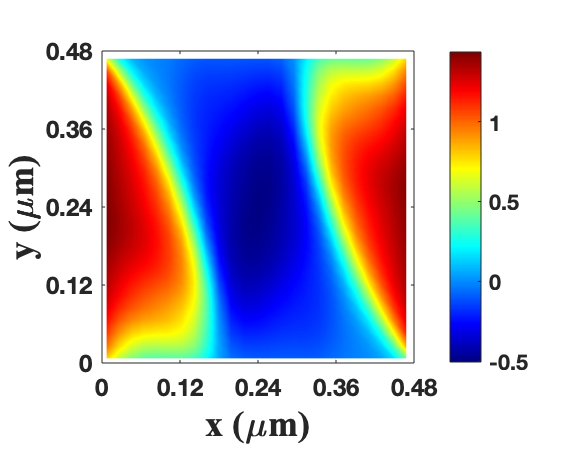}}
	\subfloat{\label{BDF1_alpha_1_ang_v1}\includegraphics[width=1.3in]{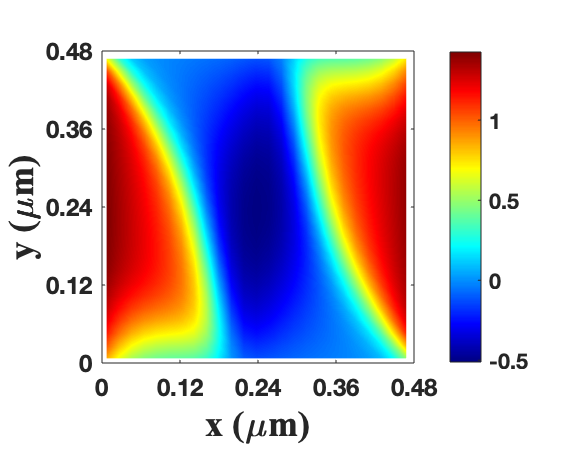}}
		\hspace{0.1in}
		\subfloat{\label{BDF1_alpha_5_ang_v1}\includegraphics[width=1.3in]{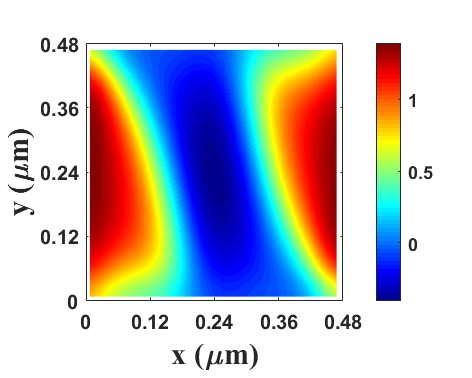}}
		\subfloat{\label{BDF1_alpha_10_ang_v1}\includegraphics[width=1.3in]{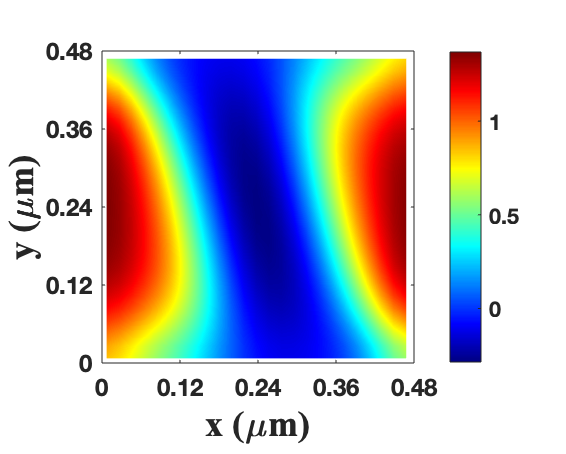}}
		\subfloat{\label{BDF1_alpha_40_ang_v1}\includegraphics[width=1.3in]{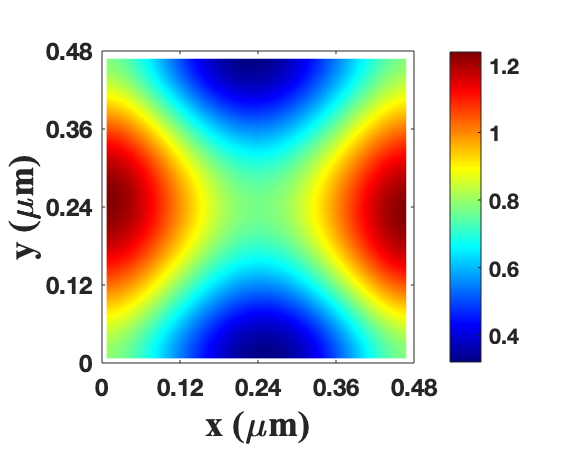}}
			\subfloat{\label{BDF1_alpha_100_ang_v1}\includegraphics[width=1.3in]{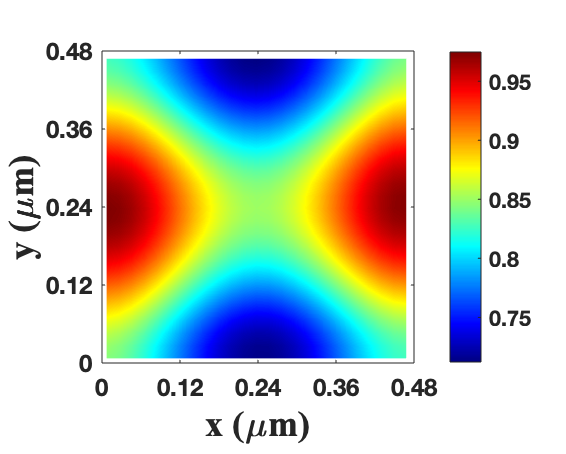}}
	\caption{Stable structures in the absence of magnetic field at $2\,$ns. The color denotes the angle between the first two components of the magnetization vector. Top two rows: Proposed method; Middle two rows: BDF2; Bottom two rows: BDF1. From left to right: $\alpha=0,0.01,0.1,1,5,10,40,100$. $dt=0.1\;ps$. }\label{BDF2_GSPM_alpha_v1}
\end{figure}

To evaluate the numerical stability of these three methods in practical micromagnetic simulations involving arbitrary damping parameters, a systematic set of numerical experiments was designed. The simulation model adopts a ferromagnetic thin film with dimensions of $480\times480\times20\,\textrm{nm}^3$, which is discretized using a computational grid of $100\times100\times4$ points—this grid resolution balances the need for capturing fine magnetic structures and maintaining computational efficiency. Two distinct temporal step-sizes, $k=1\,\text{ps}$ and $k=0.1\,\text{ps}$, were employed to investigate the influence of time discretization on stability performance.
The initial magnetization configuration was set to a uniform state aligned along the $x$-direction, and the external magnetic field was configured to zero to eliminate external driving effects, thereby isolating the impact of damping parameters on system stability. A broad range of damping parameter values was tested to cover both weak, moderate, and extremely strong dissipation scenarios: $\alpha=0,0.01,0.1,1,5,10,40,100$. The magnetization profiles obtained from these simulations, which confirm the stable evolution of the magnetic system under all tested conditions, are presented in \cref{BDF2_GSPM_alpha} and \cref{BDF2_GSPM_alpha_v1}. Based on the comprehensive simulation results, the following key observations can be drawn.
\begin{itemize}
	\item The proposed method is unstable for very large damping parameters, while BDF2 can get reasonable configurations and BDF1 get different stable results;
	\item All three methods are stable for moderately large $\alpha$;
	\item The proposed method is the only one that is unstable for small $\alpha$.
	\item The results are stable for smaller time step $k$.
\end{itemize}
In fact, a preliminary theoretical analysis conducted for the proposed numerical method has yielded insightful results regarding its convergence performance. Specifically, this theoretical framework demonstrates that an optimal-order convergence estimate of the method can be rigorously justified under the condition that the Gilbert damping parameter satisfies \(\alpha>\sqrt{2}/2\). This theoretical threshold provides a clear mathematical basis for understanding the method’s convergence behavior in scenarios with moderate-to-strong damping.
Complementing the theoretical findings, an extensive set of numerical experiments has been carried out to validate the method’s practical stability characteristics. These experiments, covering a wide range of damping parameter values and simulation configurations, consistently imply that a more relaxed condition \(\alpha>0.1\) is sufficient to guarantee robust numerical stability in practical micromagnetic computations. This discrepancy between the theoretical convergence threshold and the practical stability criterion highlights the conservatism inherent in the preliminary theoretical analysis, while also confirming the method’s favorable stability performance in real-world simulation scenarios.

\begin{figure}[htbp]
	\centering
	\subfloat[Proposed]{\label{energy_BDF3}\includegraphics[width=1.8in]{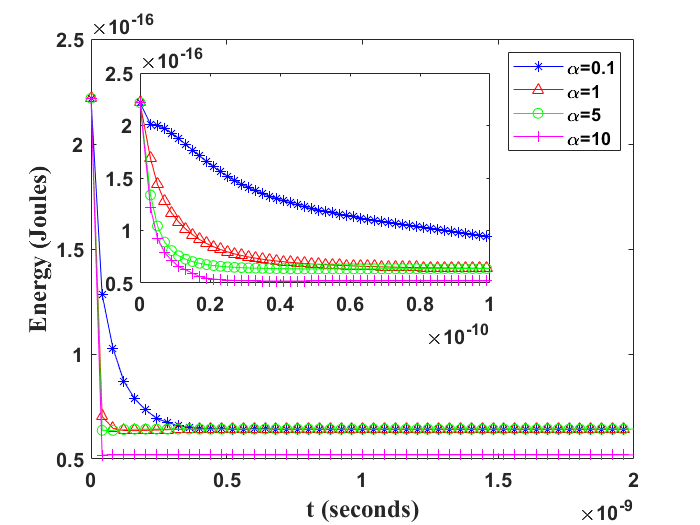}}
	\subfloat[BDF1]{\label{energy_BDF1}\includegraphics[width=1.8in]{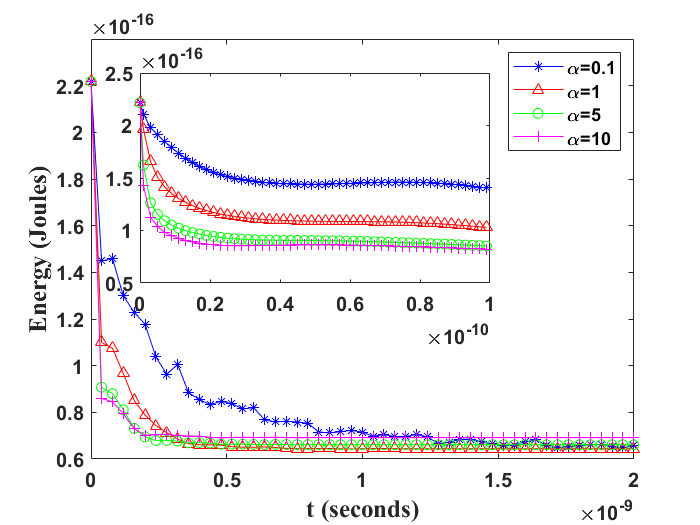}}
	\subfloat[BDF2]{\label{energy_BDF2}\includegraphics[width=1.8in]{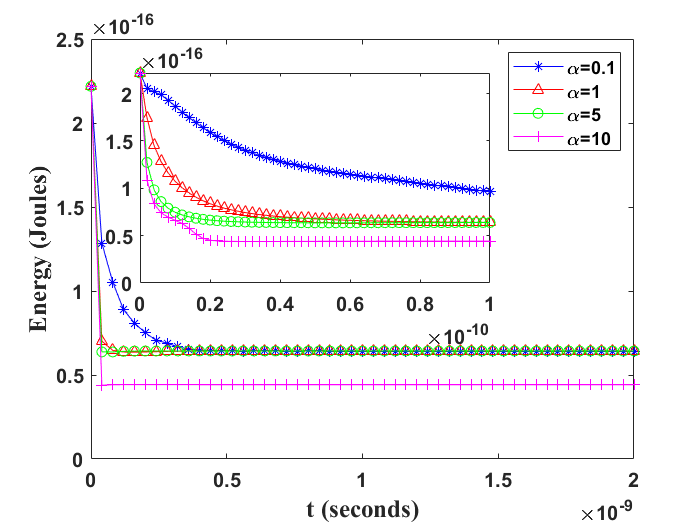}}
	\caption{Energy evolution curves of three numerical methods, with different damping constants, $\alpha=0.1,1,5,10$, up to $t=2\,$ns  in the absence of external magnetic field. Left: Proposed numerical method; Middle: BDF1; Right: BDF2. One common feature is that the energy dissipation rate is faster for larger $\alpha$, which is physically reasonable.}\label{energy_decay}
\end{figure}

Under the identical simulation setup detailed in the preceding sections, we conducted a comparative investigation into the energy dissipation characteristics of three numerical methods: the proposed scheme, the second-order backward differentiation formula (BDF2), and the first-order backward differentiation formula (BDF1). The magnetic system reaches a stable state by \(t=2\,\text{ns}\) across all tested configurations, and the total magnetic energy throughout the simulation process is calculated using the energy formulation given in \eqref{LL-Energy}.
The energy evolution curves corresponding to the three methods, each tested under four distinct damping parameter values (\(\alpha=0.1,1,5,10\)), are presented in \cref{energy_decay}. A salient and consistent feature observed across all three numerical schemes is that the rate of energy dissipation accelerates with an increase in the Gilbert damping parameter \(\alpha\). Specifically, simulations with larger \(\alpha\) values exhibit a more rapid decline in total magnetic energy until the system stabilizes, which aligns with the physical intuition that higher damping facilitates faster energy dissipation.
To validate these numerical observations, a parallel theoretical derivation was performed for the LLG equation. This derivation, as summarized in \eqref{c1-large}, explicitly demonstrates that the energy dissipation rate inherent to the LLG equation is a function of \(\alpha\), with a direct positive correlation: a larger \(\alpha\) value inherently leads to a faster rate of energy dissipation. Consequently, the energy evolution trends captured by the BDF1, BDF2, and proposed numerical methods all show excellent agreement with the conclusions derived from the theoretical analysis of the LLG equation, further confirming the physical consistency and reliability of the three numerical schemes.

\begin{figure}[htbp]
	\centering
	\subfloat[$\alpha=0.1$]{\label{alpha_0dot1}\includegraphics[width=2.5in]{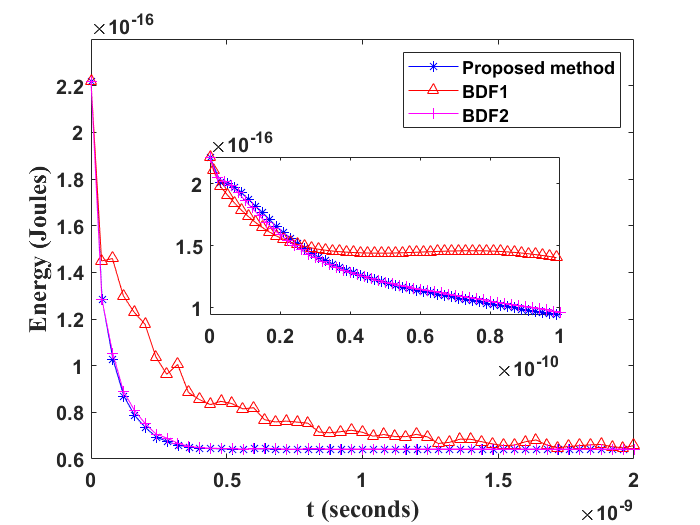}}
	\subfloat[$\alpha=1$]{\label{alpha_1}\includegraphics[width=2.5in]{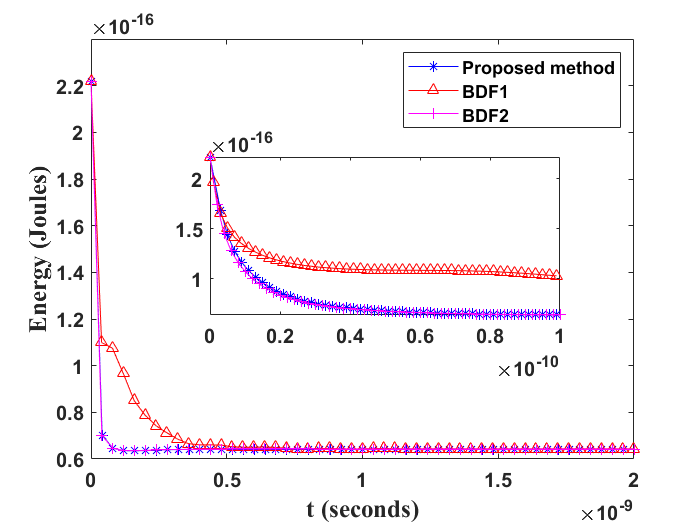}}
	\hspace{0.1in} 
	\subfloat[$\alpha=5$]{\label{alpha_5}\includegraphics[width=2.5in]{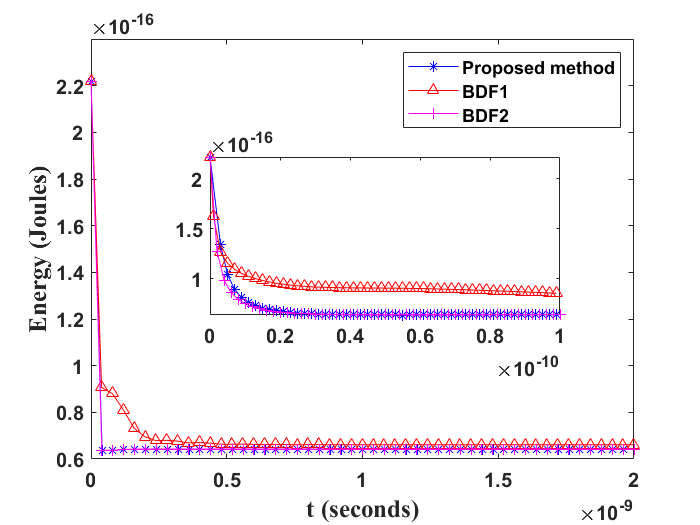}}
	\subfloat[$\alpha=10$]{\label{alpha_10}\includegraphics[width=2.5in]{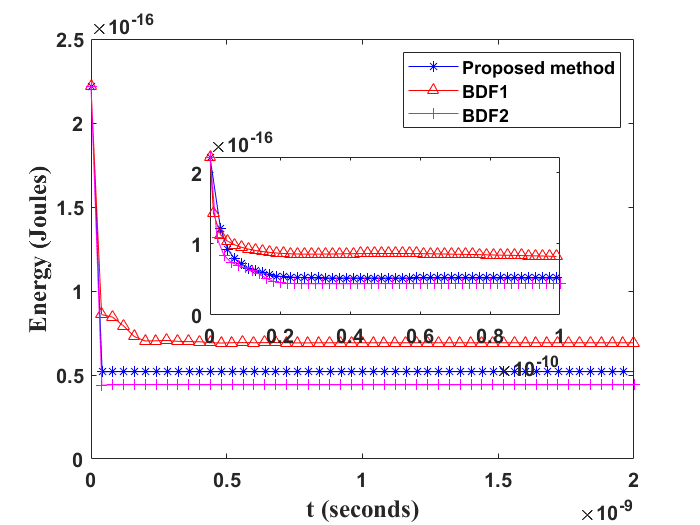}}
	\caption{Energy evolution curves in terms of time, for the numerical results created by three numerical methods up to $t=2\,$ns in the absence of external magnetic field for (a) $\alpha=0.1$, (b) $\alpha=1$, (c) $\alpha=5$, and (d) $\alpha=10$. The energy dissipation pattern of the proposed method is consistent with the BDF2 method for $\alpha=0.1,1,5$, and inconsistent with other two methods for $\alpha=10$. The rate of energy dissipation of the BDF1 method is slower than that of other methods. The energy of BDF2 and the proposed method when they reach a steady state is lower than that of BDF1 when it reaches a steady state.}\label{energy_decay_alpha}
\end{figure}

To further elaborate on the energy dissipation characteristics of the numerical schemes, we adopted the same sequence of Gilbert damping parameter values (\(\alpha\)) as employed in the preceding analysis. Corresponding energy evolution curves, tracking the total magnetic energy over time up to \(T=2\,\text{ns}\) (the time at which the system reaches a stable state), are presented in \cref{energy_decay_alpha}.
A detailed comparison of these curves reveals several notable patterns. For damping parameter values of \(\alpha=0.1,1,5\), the energy dissipation pattern exhibited by the proposed method is consistent with that of the BDF2 scheme, indicating a high degree of agreement in energy evolution behavior under weak-to-moderate damping conditions. However, this consistency breaks down when \(\alpha=10\) (a strong damping scenario): the energy dissipation pattern of the proposed method deviates from both the BDF1 and BDF2 schemes.
Among the three methods, the BDF1 scheme demonstrates the slowest rate of energy dissipation across all tested \(\alpha\) values. Additionally, a key quantitative difference emerges in the steady-state energy levels: the total magnetic energy of the system when reaching stability via the BDF2 and proposed methods is consistently lower than the steady-state energy achieved with the BDF1 method. This discrepancy in steady-state energy further underscores the varying capabilities of the three numerical schemes in capturing the energy dissipation physics inherent to the LLG equation, particularly under strong damping conditions.

\subsection{Magnetic Domain wall motion}
\begin{figure}[htbp]
	\centering
	\subfloat[Magnetization for initial state]{\label{NeelWall_initial_mag}\includegraphics[width=2.8in]{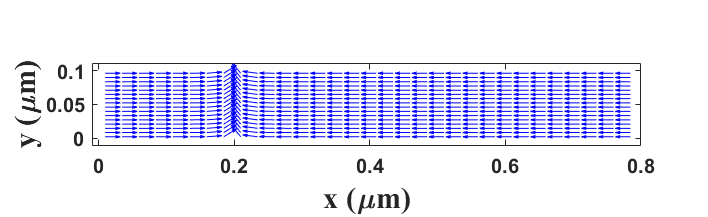}}
	\subfloat[Magnetization with $\alpha=0.1$  at $1.6\,$ns]{\label{NeelWall_alpha_0dot1_mag}\includegraphics[width=2.8in]{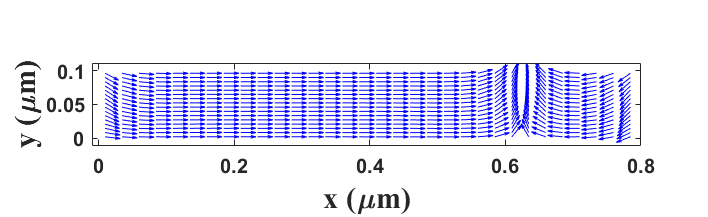}}
	\hspace{0.1in}
	\subfloat[Magnetization with $\alpha=0.4$ at $1.6\,$ns]{\label{NeelWall_alpha_0dot4_mag}\includegraphics[width=2.8in]{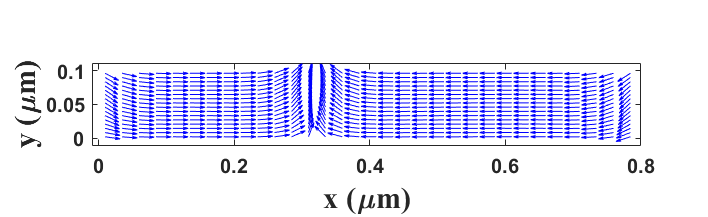}}
	\subfloat[Magnetization with $\alpha=0.8$ at $1.6\,$ns]{\label{NeelWall_alpha_0dot8_mag}\includegraphics[width=2.8in]{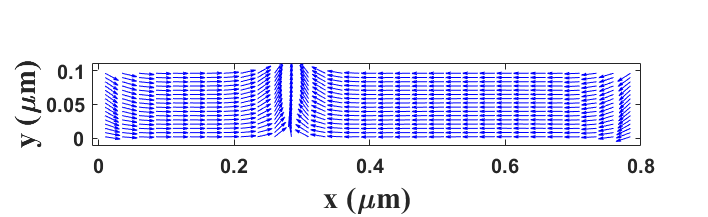}}
	\hspace{0.1in}
	\subfloat[Magnetization with $\alpha=1$ at $1.6\,$ns]{\label{NeelWall_alpha_1_mag}\includegraphics[width=2.8in]{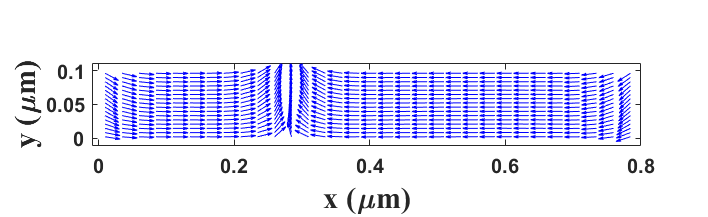}}
	\subfloat[Magnetization with $\alpha=2$ at $1.6\,$ns]{\label{NeelWall_alpha_2_mag}\includegraphics[width=2.8in]{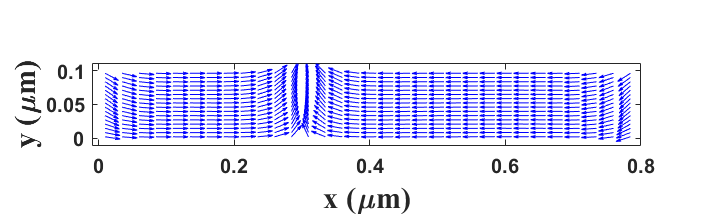}}
	\hspace{0.1in}
	\subfloat[Magnetization with $\alpha=3$ at $1.6\,$ns]{\label{NeelWall_alpha_3_mag}\includegraphics[width=2.8in]{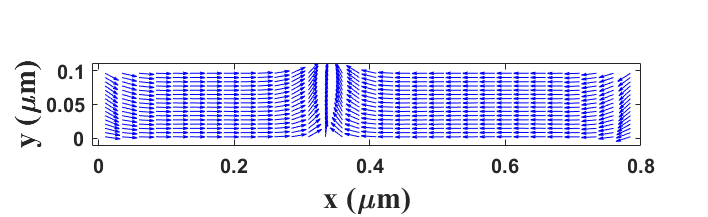}}
	\subfloat[Magnetization with $\alpha=5$ at $1.6\,$ns]{\label{NeelWall_alpha_5_mag}\includegraphics[width=2.8in]{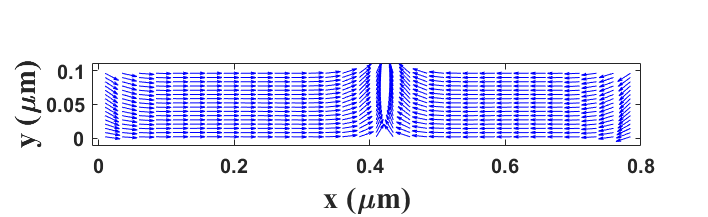}}
	\caption{Magnetization profiles of Ne\'{e}l wall motion in the presence of a magnetic field $\h_e=5\,$mT, with $\alpha = 0.1,0.4,0.8,1,2,3,5$ at $1.6\,$ns for the proposed numerical method. The in-plane arrow denotes the first two components of the magnetization vector. The wall moves slower for larger values of $\alpha$ (when $0<\alpha<1$) and faster for larger values of $\alpha$ (when $\alpha>1$)  and its velocity depends quadratically on $\alpha$. The velocity roughly achieves its minimum value around $\alpha=1$.}\label{NeelWall_alpha_2ns}
\end{figure}

\begin{figure}[htbp]
	\centering
	\subfloat{\label{velocity_alpha_varied_He}\includegraphics[width=2.8in]{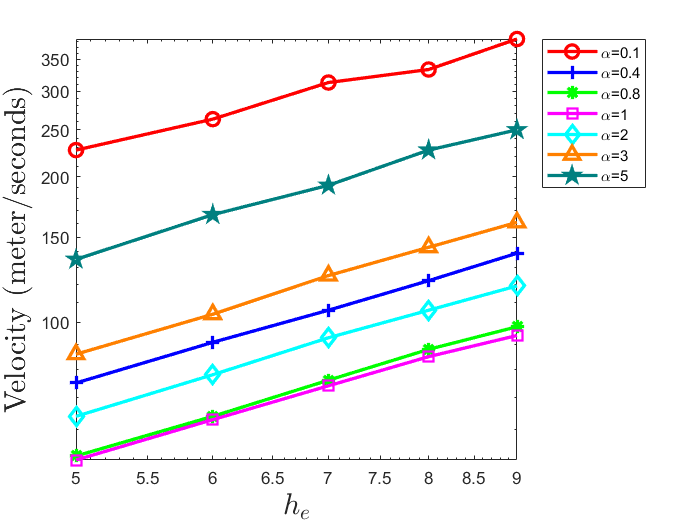}}
	\subfloat{\label{velocity_He_fixed_alpha}\includegraphics[width=2.8in]{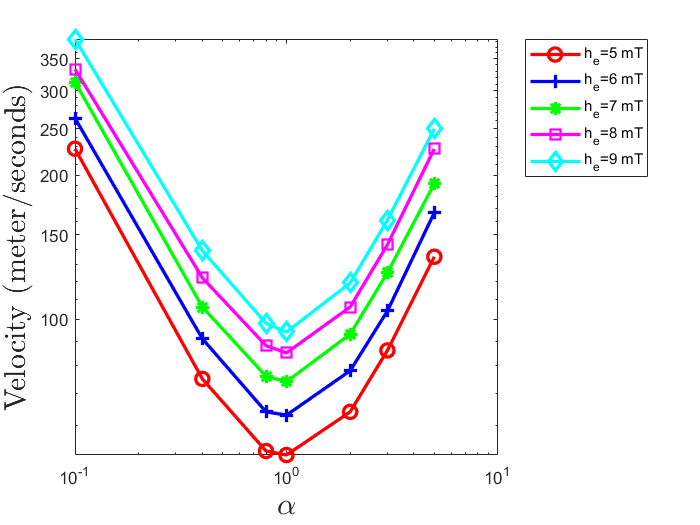}}
	\caption{Linear dependence of the wall velocity with respect to the external magnetic field $\h_e$ (left); Quadratic dependence of the wall velocity with respect to the damping parameter  $\alpha$ (right).}\label{velocity_alpha_He}
\end{figure}

A Neél domain wall was established as the initial magnetic state within a ferromagnetic nanostrip characterized by size of \(800\times100\times4\,\textrm{nm}^3\). To ensure sufficient spatial resolution for capturing domain wall features while maintaining computational feasibility, the nanostrip was discretized using a structured grid comprising \(128\times64\times4\) nodes. Subsequent to the initialization process, an external magnetic field with a magnitude of \(\h_e=5\,\text{mT}\) was applied along the positive \(x\)-direction to induce domain wall motion. Micromagnetic simulations of the domain wall dynamics were then performed over a time interval of up to \(1.6\,\text{ns}\), with the Gilbert damping parameter \(\alpha\) systematically varied across seven distinct values: $0.1, 0.4, 0.8, 1, 2, 3,$ and $5$. The magnetization spatial distributions obtained from these simulations, which directly reflect the evolution of the Neél wall under different damping conditions, are presented in \cref{NeelWall_alpha_2ns}.
Qualitative assessment of the simulation outcomes reveals a consistent trend: for a fixed value of the damping parameter \(\alpha\), the propagation velocity of the Neél wall increases monotonically with the strength of the external magnetic field \(\h_e\). This qualitative observation is further substantiated by quantitative data analysis, which confirms that the relationship between domain wall velocity and \(\h_e\) exhibits a strict linear characteristic. This linear dependence is clearly illustrated in the left panel of \cref{velocity_alpha_He}, where the velocity data points align closely with the lines derived from linear fitting. To quantify the sensitivity of domain wall velocity to \(\h_e\) under different damping levels, the slopes of these linear fits were calculated using the least-squares method for each \(\alpha\) setting. These slope values, which serve as key metrics for characterizing the field-response behavior of the domain wall, are compiled in \cref{tab-3}. On the other hand, when we fix \(\h_e\), the domain wall velocity exhibits a quadratic function trend with respect to the value of \(\alpha\). Specifically, the velocity reaches a minimum value around \(\alpha = 1\): when \(0 < \alpha < 1\), the velocity shows a decreasing trend; when \(\alpha > 1\), the velocity shows an increasing trend. Indeed, in our previous work \cite{xie2025schemeB}, we have proposed an alternative format, in which it is also verified that for a relatively large dissipation coefficient, the domain wall velocity at this point increases with both $\alpha$ and $\h_e$, and such an increase shows a linear dependence on both parameters.

\begin{table}[htbp]
	\centering
	{\caption{Linear dependence of the wall velocity $V$ with respect to the external magnetic field $\h_e$ and nearly quadratic dependence of the wall velocity with respect to the damping parameter $\alpha$. } \label{tab-3} }{
		\begin{tabular}{c|c|c|c|c|c|c}
			\hline 
			\diagbox{$\alpha$}{$V$ (m/s)}{$\h_e(\textrm{mT})$}&5&6&7 &8 & 9 & Slope\\
			\hline
			0.1& 227 &263  &313  &333  &385 &0.883 \\
			0.4& 75 &91 &106  &122  &139 & 1.043\\
			0.8&53 & 64&76  &88  & 98& 1.059\\
			1& 52 &63 &74  &85  &94 & 1.016 \\
			2& 64& 78& 93 &106  &119 &1.060 \\
			3& 86 & 104 &125  &143  &161 & 1.076 \\
			5& 135 & 167 &192  &227  & 250&1.054 \\ 
				\hline 
			Slope & -&-  &-  &-  &- &  --\\
			\hline 
		\end{tabular}
	}
\end{table}

\section{Conclusions}
\label{sec:conclusions}

In the present work, a third-order accurate numerical method is proposed for micromagnetic simulations. To facilitate numerical implementation, the governing system is reorganized such that the damping term is re-expressed as a harmonic mapping flow. The core of this numerical scheme lies in three key components: third-order backward differentiation formula (BDF3) approximation for the temporal derivative, implicit discretization of the constant-coefficient diffusion term, semi-implicit treatment of the cross-product nonlinear term, and fully explicit extrapolation for the remaining terms.
Owing to its third-order accuracy, the proposed method exhibits significantly higher efficiency compared to existing lower-order semi-implicit schemes. Comprehensive numerical results from both one-dimensional (1D) and three-dimensional (3D) simulations are presented to validate the accuracy and efficiency of the developed method. Furthermore, micromagnetic simulations employing the proposed method yield physically consistent microstructures and successfully capture the key dependencies of domain wall velocity: a linear relationship with the external magnetic field and a quadratic relationship with the damping parameters.

\section*{Acknowledgments}
This work is supported in part by the grants NSF DMS-2012669 (C.~Wang), and Jiangsu Science
and Technology Programme-Fundamental Research Plan Fund (BK20250468), Research and Development Fund
of XJTLU (RDF-24-01-015) (C. Xie).

\bibliographystyle{amsplain}
\bibliography{references}

\end{document}